\documentclass[aps,prc,reprint,amssymb,amsmath,superscriptaddress, floatfix,10pt]{revtex4-2}
\usepackage{mathptmx}
\usepackage{graphicx}
\usepackage{color}  
\makeindex
\newcommand{\textblue}{\textcolor[rgb]{0.00,0.07,1.00}}

\begin{document}

\title{The compressible liquid drop model with curvature included algebraically for the pasta phases in neutron stars}

\author{Dmitry Kobyakov}
\email{dmitry.kobyakov.nsp@gmail.com}
\affiliation{Independent Researcher, Saint-Petersburg, Russia}

\begin{abstract}
In the compressible liquid drop model (CLDM), the curvature energy term has been included in many works in the literature, and its effects on the presence of pasta phases have also been extensively investigated.
However, the analytical part of the inclusion used to cease once a 4-order algebraic equation for the optimized size of the nucleus was written.
In this paper, an optimization is suggested to the analytical part of the solution of the CLDM equations.
Solving analytically the 4-order algebraic equation proves to be helpful for a straightforward identification of the denied parameter ranges in which the CLDM has no real-root solutions.
Using a particular nuclear force derived from the chiral effective field theory, I investigate effects on the ground state of the nucleus curvature. 
Surprisingly, the inclusion has lead to a significant modification of the range of baryon density in which the bubble phases are found, disfavoring the bubble phases from the ground state.
This shows a strong difference between the predictions of the CLDM with the nucleus size obtained when including the curvature term and that obtained from the $w_\mathrm{s} = 2w_{\mathrm{C+L}}$ approximation.
The importance of the curvature energy in the CLDM is linked with its ability to change qualitatively the ground state.
This paper presents an analytical extension to the CLDM solution, which is ready to be used with any other nuclear model in the future, in order to optimize the inclusion of the nuclear curvature energy, and is not limited to one nuclear model.
\end{abstract}

\maketitle

\section{Introduction}
Astrophysical extrapolations of nuclear theory to neutron star matter have received a significant attention:
The theory valid for description and prediction of physical properties of matter inside the atomic nucleus is evolved towards the regime of subnuclear and uniform nuclear densities, which are expected to exist inside neutron stars \cite{BBP1971,BulgacMagierski2001,ChenPiekarewicz2014,YeEtal2025}.
Various astrophysical interests of studying the pasta phases in neutron stars are well-known.

The extrapolation however evolves also the basic physical observables in the problem.
While in atomic nucleus those are the mass number $A$, the charge number $Z$ and the binding energy per baryon $E_B/A$, in neutron star matter the basic observables are the stellar mass and the stellar radius.
These properties put robust constrains on the equation of state of neutron star matter, which includes the information mainly on the uniform nuclear matter in the core and also on the ordinary nuclei near the edge of the star and on the exotic matter in the transition region in the inner crust, where the nuclei are expected to be nonspherical in their ground state.
Clearly, the nuclei of the inner crust cannot be studied in the laboratory and the information on their $A$, $Z$ and $E_B/A$ cannot be directly observed.
In this case, the basic quantities are the characteristic properties of the equation of state of symmetric uniform nuclear matter, such as the saturation nuclear density $n_0$, the binding energy per baryon $B$, the incompressibility $K$, the nuclear symmetry energy $S$ and its slope $L$.
It is worth mentioning that in the relation of the Skyrme theory applied to uniform nuclear matter and that applied to atomic nucleus, there is a robust correlation between the saturation nuclear density $n_0$ and the charge nucleus radius, so is between $L$ and the neutron skins \cite{ReinhardNazarewicz2016}.

The equation of state of neutron star matter can be studied by astronomical observations that target specific layers of the star.
For a recent account, see \cite{XiaEtal2021} and references therein.
For instance, in low mass X ray binaries, the mass shedding from the companion star onto the neutron star generates quasi-periodic cycles of heating and cooling, which is a very useful probe for thermodynamic properties of the inner crust matter.
Observations of highly magnetized neutron stars (magnetars) suggest the existence of yet unknown physical mechanisms that trigger and drive x-ray and gamma-ray bursts, which are associated with the magnetohydrodynamics of neutron star interior including the inner crust.
Yet hydrodynamics and elasticity of neutron stars is believed to underlie the phenomenon of rotational glitches in pulsars.
In addition, the elastic properties of the crust are relevant in theory of timing features in the aftermath of the magnetar bursts and in theory of the tidal deformability during the gravitational wave emission in neutron star mergers.
Thus, there is a robust connection between the observations and the equation of state of the inner crust matter.

The inner crust contains physics of a nucleus at mean baryon densities in the range of subnuclear values.
This implies that nuclei are expected to become extremely rich in neutrons and to possess positive pressure.
The latter is balanced by the pure neutron matter outside the nucleus, while the electron background ensures the chemical equilibrium of neutrons and protons.
In this case, the astrophysical extrapolation of theory of atomic nucleus supplemented by theory of pure neutron matter is appropriate.

Understanding the multitude of macroscopic dynamic phenomena requires to possess not only the equation of state of neutron star matter, but also theoretical frameworks, which might be conveniently used to study the relevant macroscopic dynamical processes.
One of the most useful frameworks is the compressible liquid drop model (CLDM) introduced in \cite{BBP1971}.
The CLDM originates from the liquid drop model (LDM) originally developed for the atomic nucleus and contains all essential ingredients of nuclear physics in the inner crust, paving the way to qualitative understanding of the matter, which is the reason why the CLDM has remained a relevant framework for over six decades.

In studies of the atomic nucleus, the LDM has provided pioneering predictions and has remained a useful approach, see \cite{MorettoEtal2012,RocaMazaEtal2013,PomorskiXiao2025} and references therein.
In the basic concept, the nucleons are assumed to be incompressible and spherical particles sticking to one another when touch, forming a nucleus with volume $V_N$ and radius $r_N=(3/4\pi)V_N^{1/3}$, thus, $ r_N\propto A^{-1/3}$, where $A$ is the atomic number.
The starting point of the LDM is the Bethe-Weizs$\ddot{\mathrm{a}}$cker mass formula for the binding energy $E_B$ of nucleus \cite{Kirson2008}.
For nuclei with equal number of protons and neutrons, it reduces to the so-called leptodermous expansion \cite{MorettoEtal2012}:
\begin{equation}\label{def_EB}
E_{\mathrm{B}}=-a_vA+a_sA^{2/3}+a_rA^{1/3}+\cdots,
\end{equation}
where the numerical constants are related to the bulk energy per nucleon ($a_v$), the plain surface tension ($a_s$) and the surface curvature tension ($a_r$) of the nucleus \cite{MorettoEtal2012}.
By fitting $E_{\mathrm{B}}$ to experimentally measured nuclear masses, one finds $a_v,\,a_s,\,a_r$ \cite{Kirson2008,LiuEtal2011,MorettoEtal2012,CherevkoEtal2014,CherevkoEtal2015,Yuan2016,JodonEtal2016,KolomietzEtal2017,CauchoisEtal2018,SagunEtal2019,CaoEtal2022,DutraLourencoMenezes2025}.

As a matter of fact, a freedom remains to assume whether $a_v$ is equal to $a_s$ and whether the information provided by $a_r$ is absorbed into $a_s$ \cite{MorettoEtal2012}.
The physical motivation is that the surface energy is in a sense a partial lack of the volume energy of the nucleon interaction near the edge of the nucleus.
Additional terms and modifications to Eq. (\ref{def_EB}) have been suggested aimed at better fits to the nuclear data \cite{MyersSwiatecki1966,PomorskiXiao2025}.
Here I avoid including the corrections related to the concepts of the neutron skin or halo, the baryon pairing and the correlations in the structure of the nuclear energy levels because of their smallness as compared to the bulk and the surface properties.
Those corrections inserted into the LDM provide a hybrid approach, which remains to date the model of excellence for more sophisticated approaches such as Hartree-Fock-Bogoliubov \cite{MorettoEtal2012}.

Analogously to the aim of the Hartree-Fock-Bogoliubov approaches in physics of the atomic nucleus, in the context of neutron star matter those are aimed at pushing theory to the next quantitative level of particularisation of predictions, as compared to more qualitative description provided by the CLDM.
The crucial difference is that the connection of theory with observations of neutron stars is qualitatively more complex as compared to observations of the atomic nucleus.
Theoretical predictions relevant to neutron star matter are contaminated with significantly larger uncertainties \cite{BulgacMagierski2001,MagierskiHeenen2002,ChenPiekarewicz2014,Tews2017}.
The latter are generated by incomplete understanding of the nucleon force, by indirect nature of observations of physical mechanisms predicted by  theory of the microscopic structure of neutron star matter and by thermodynamic fluctuations, which are far from negligible in the real matter.
This explains why going to the next level of particularisation alone does not immediately improve understanding of the real material and why approaching the problem from various directions is useful.
This motivates further studies in the framework of the CLDM.

Determination of the equilibrium parameters of nuclei in the CLDM is a variational problem.
Its modern formulation has been presented in \cite{WatanabeEtAl2000} and is based on the total energy of the system, which is assumed to consist of equal unit cells.
The energy of a unit cell is
\begin{equation}\label{def_Etot_I}
E_{\mathrm{tot}}=E_{\mathrm{B}}+E_{0}+E_{\mathrm{C+L}}+E_{\mathrm{e}}+E_{\mathrm{no}},
\end{equation}
where $E_{0}$ is the rest-mass energy of baryons, $E_{\mathrm{C+L}}$ is the Coulomb energy of protons and electrons, $E_{\mathrm{e}}$ is the mechanical energy of the electrons and $E_{\mathrm{no}}$ is the energy of pure neutron liquid outside the nucleus.
In this case, the nucleus binding energy is
\begin{equation}\label{def_EB_I}
E_{\mathrm{B}}=V_N\left(n\varepsilon + \frac{d}{r_N}\sigma_s + \frac{d(d-1)}{r_N^2}\sigma_c\right),
\end{equation}
where $\varepsilon=\varepsilon(n,x)$ is the energy per baryon in uniform nuclear matter, $d$ is the number of spatial dimensions along which the nucleus has the surface, $x$ is the proton fraction inside the nucleus, $\sigma_s=\sigma_s(x)$ is the surface tension and $\sigma_c=\sigma_c(x)$ is the curvature tension, which are functions of $x$ only.
For spherical nuclei ($d=3$), the binding energy in Eq. (\ref{def_EB_I}) has a similar form as the original Eq. (\ref{def_EB}).
However, while $E_{\mathrm{B}}$ is directly observable in the LDM, but it is not in the CLDM.

It is interesting to notice in Eq. (\ref{def_EB_I}), that while $\varepsilon$ is a function of both the baryon density and the proton fraction, but $\sigma_s$ and $\sigma_c$ are functions of the proton fraction only \cite{NewtonGearheartLi2013}.
This is a consequence of the perturbative character of inclusion of the surface energy into the theory by separating it from the bulk (volume) part.
Specifically, $\varepsilon$ is calculated in uniform nuclear matter, where pressure is a free parameter and the beta equilibrium is not required in such calculations.
Since the uniform nuclear matter is completely determined by just two properties, $n$ and $x$, therefore those remain free.
However, the calculation of $\sigma_s$ and $\sigma_c$ is based on a model problem with two components: Semi-infinite nuclear matter (extending to, say, minus infinity of a coordinate axis) in equilibrium with semi-infinite pure neutron liquid (extending to plus infinity of the same axis), with a smooth transition between them (in vicinity of zero of the axis).
Such two-component system is completely described by $n$, $x$ and the number density $n_{no}$ of neutrons in the pure neutron liquid \cite{RavenhallPethickLattimer1983,CentelesEstalVinas1998,DouchinHaenselMeyer2000}.
The equilibrium requires that the pressures in the bulks of the both components are equal and that the corresponding neutron chemical potentials in the both components are equal.
These two requirements yield two equations, which fix two variables of the set $n$, $x$, $n_{no}$ and, thus, leaving only one of those free, which we choose to be the proton fraction $x$.

The parameters of the CLDM such as $\varepsilon$, $\sigma_s$ and $\sigma_c$ are calculated from the microscopic physics involving the nucleon-nucleon force.
There are many calculations of the uniform nuclear matter part $\varepsilon$, see \cite{BBP1971,HebelerEtal2013} and references therein.
Likewise, the surface tension coefficients $\sigma_s$ and $\sigma_c$ were calculated in many works \cite{BBP1971,RavenhallBennettPethick1972,NegeleVautherin1973,BennettRavenhall1974,RavenhallPethickLattimer1983,StockerFarine1985,Farine1988,DurandSchuckVinas1993,CentellesVinasSchuck1996,CentelesEstalVinas1998,EstalCentellesVinas1999,DouchinHaenselMeyer2000,NakazatoIidaOyamatsu2011,NewtonGearheartLi2013,LimHolt2017}.
The consistency requires that the same underlying nucleon force parametrization is used in calculation of $\varepsilon$, $\sigma_s$ and $\sigma_c$.
However, often in the earlier work the uniform part and the surface parts were calculated independently \cite{NewtonGearheartLi2013,LimHolt2017}.
Let us consider three chosen issues remaining in theoretical incorporation of the surface effects into the CLDM.

(i) In the modern literature the quantity $\sigma_c$ is often neither included in calculations at all, nor it is absorbed by $\sigma_s$ \cite{WatanabeEtAl2000,NakazatoOyamatsuYamada2009,LimHolt2017}.
An example of exception is provided by \cite{NakazatoIidaOyamatsu2011}, where $\sigma_c$ was taken into account, but an even larger approximation than neglecting $\sigma_c$ was made, since the values of $n$, $n_{no}$ and $x$ were set by hand, as was explicitly stated in the last paragraph of Sec. II in \cite{NakazatoIidaOyamatsu2011}.
However, earlier a considerable sensitivity of the pasta phase transitions on $\sigma_c$ had been pointed out \cite{PethickRavenhallLattimer1983}.
Indeed, calculations reported in this paper reveal that inclusion of $\sigma_c$ alters the appearance of the bubble phases significantly.

(ii) The surface tension $\sigma_s$ was calculated in \cite{LimHolt2017}, see also \cite{NewtonGearheartLi2013}, by fitting of an empirical (trial) function (see Eq. (9) in \cite{LimHolt2017}).
This approach is very similar to the one used in \cite{BBP1971,WatanabeEtAl2000}.
However, \emph{a priori} there is no reason for $\sigma_s$ to obey the dependence on the proton fraction $x$ inside the nucleus assumed by the trial fitting function.
Indeed, the data on $\sigma_s(x)$ obtained directly from numerical calculations is notably different from the trial fitting function.
To reveal this difference, in this paper I extract the surface contributions to the CLDM from the Skyrme model in Thomas-Fermi approximation \cite{RavenhallBennettPethick1972,BrackGuetHakanson1985} of semi-infinite nuclear matter supported by semi-infinite pure neutron liquid following the method described in Refs. \cite{CentelesEstalVinas1998,DouchinHaenselMeyer2000}.
In these calculations, the consistency of the bulk ($\varepsilon$) and the surface ($\sigma_s$ and $\sigma_c$) parts is guaranteed by using the same underlying nucleon force.

(iii) The variable $r_N$, to the best of my knowledge, was computed approximately in Eq. (6) of \cite{PethickRavenhallLattimer1983} and in the following literature.
The explicit analytical formula for the roots of a fourth-order polynomial that determines $r_N$ was not used.
In terms of Eq. (\ref{def_EB_I}), the approximation method assumes that $\sigma_s \gg {(d-1)}\sigma_c/r_N$, although this is actually not the case (see Fig. 4 and Table III below).
Here, I reveal the analytical algebraic solution for $r_N$, which allows to exclude the standard numerical approximations in calculation of $r_N$, thereby effectively reducing the number of variables in the numerical formulation of the variational problem, accounting for the curvature exactly and, thus, advancing the implementation of solution of the CLDM.
Even more importantly, for a given set of the parameters in the unit cell, the mean baryon number density $n_b$, the number density in the neutron matter $n_{no}$ and the number density $n$ and the proton fraction $x$ in the nuclear matter, the algebraic solution to $r_N$ (or $r_B$ for bubbles) allows a straightforward check to whether it is real-valued or complex-valued and therefore to whether the CLDM equilibrium exists or does not exist.

In this paper I will focus on advancement of the mathematical implementation of the CLDM, on reducing the approximations in numerical evaluation by carefully including the curvature into the nuclear binding energy and on the consistency between the calculation of the bulk and the surface nuclear properties.
The surface tension will be calculated explicitly using the Skyrme nuclear model in the Thomas-Fermi approximation based on the nucleon force derived from the chiral effective field theory.
I will show that this contribution although represents a correction to the commonly used version of the CLDM, but its effect on the inner crust structure predicted by the CLDM is qualitative: Its inclusion leads to a significant narrowing of the density range in which the bubble phases exist.
Also, I will calculate the energy per baryon in the standard phases of the nonspherical nuclei of the inner crust and their separation from the ground state energy and compare the differences with the expected thermal energy per baryon for three different temperatures.

The paper is organized as following.
In Sec. II the CLDM is described and its basic equations are presented.
Section III is devoted to the solution scheme of the CLDM, where I also present the new analytical calculation of the nucleus radius.
Numerical evaluation of the parameters $\varepsilon$, $\sigma_s$ and $\sigma_c$ of the CLDM from the nuclear theory is explained in detail in Sec. IV.
Using these parameters, I solve numerically the CLDM in Sec. V and display the solutions in Figs. 3-6.
The astrophysical implications and the addition of this work to the existing knowledge are discussed in Sec. VI.
Section VII contains the concluding remarks.
Appendix is reserved for the essential information on the nucleon force parametrization used in this work, for the numerical results on the surface tension calculated in the Thomas-Fermi approximation, for the part of the theory relevant exclusively to the bubble phases and for a detailed algebraic derivation of $r_N$ and $r_B$.

\section{Compressible liquid drop model with curvature}
\subsection{Basic definitions}
Locally inside the nucleus, the nuclear matter is assumed uniform.
A complete set of physical quantities needed to specify the ground state consists of
the radius of nucleus $r_N$ (a scalar), the baryon number density $\tilde{n}$ and the proton fraction $\tilde{x}$ both inside nucleus (spatial functions constant inside and zero outside nucleus) and the number density of neutrons outside nucleus $\tilde{n}_{no}$ (a spatial function constant outside and zero inside the nucleus).

Let the origin of the spatial coordinates $\mathbf{r}$ be in the center of the nucleus.
Then limiting the domain of our consideration to the volume of the unit (Wigner-Seitz) cell $V_c$, $\mathbf{r}\in V_c$, we can assume
\begin{eqnarray}
 && \tilde{n}(\mathbf{r})=n\theta(r_N-|\mathbf{r}|), \\
 && \tilde{x}(\mathbf{r})=x\theta(r_N-|\mathbf{r}|), \\
 && \tilde{n}_{no}(\mathbf{r})=n_{no}\theta(|\mathbf{r}|-r_N),
\end{eqnarray}
where $\theta(y)$ is the Heaviside step function.
At {fixed} $n_b$, the quantity $V_c$ is uniquely determined by the choice of
\begin{equation}\label{VarSet1}
  \{r_N,\,\tilde{n}(\mathbf{r}),\,\tilde{x}(\mathbf{r}),\,\tilde{n}_{no}(\mathbf{r})\}.
\end{equation}
It is easy to see that the system can be described equivalently by the set of parameters
\begin{equation}\label{VarSet2}
  \{r_N,\,{x},\,{n}_{ni},\,u\},
\end{equation}
where
\begin{equation}\label{def_nni}
\tilde{n}_{ni}(\mathbf{r})={n}_{ni}\theta(r_N-|\mathbf{r}|),
\end{equation}
with ${n}_{ni}$ being the number density of neutrons inside the nucleus defined in Eq. (\ref{def_npnni}) and
\begin{equation}\label{def_u}
  u=\frac{V_N}{V_c},
\end{equation}
is the volume fraction of the nucleus in the unit cell.
The set defined in Eq. (\ref{VarSet2}) will be used in the basic formulation in this paper.

The unit cell with volume $V_c$ contains the total number of neutrons $N_n$ and protons $N_p$.
In the unit cell, $N_n$ is a sum of the number of neutrons inside and outside the nucleus,
\begin{equation}\label{def_Nss}
  N_n=N_{ni}+N_{no}.
\end{equation}
The protons are present only inside and are absent outside the nucleus.
Inside the nucleus, the number density of protons $n_p$ is
\begin{equation}\label{def_npnni}
  n_p=\frac{N_p}{V_N}=xn, \quad n_{ni}=\frac{N_{ni}}{V_N}=(1-x)n.
\end{equation}
The nucleus volume $V_N$ is given in Eq. (\ref{def_VN}).

Notice that in thermodynamic equilibrium, both $V_c$ and $V_N$ are determined by the variational equations, while the mean baryon number density $n_b$,
\begin{equation}\label{def_nb}
  n_b=\frac{N_p+N_{ni}+N_{no}}{V_c}=un+(1-u)n_{no},
\end{equation}
is a fixed input parameter.
The baryon number density inside the nucleus is
\begin{equation}\label{def_nb}
  n=\frac{N_p+N_{ni}}{V_N}
\end{equation}
and the number density of the pure neutron liquid is
\begin{equation}\label{def_nno}
  n_{no}=\frac{N_{no}}{V_c-V_N}.
\end{equation}
The proton fraction is
\begin{equation}\label{def_x}
  x=\frac{N_{p}}{N_{p}+N_{ni}}.
\end{equation}

In the bubble phases, the unit cell has an inverted structure, when the cell is filled with the nuclear matter while the latter surrounds a bubble in the cell center that is filled with the pure neutron liquid.
For convenience, the basic parameters for the bubble phases are provided in Appendix.
Correspondingly, the types of unit cell are called pasta phases \cite{CaplanHorowitz2017} and are denoted as 1N (lasagna), 2N (spaghetti), 3N and for the bubble phases 2B, 3B.

In case of 3N and 3B, the nuclear matter and the pure neutron liquid are separated by a spherical interface.
Correspondingly, in case of 2N and 2B the interface is a cylindrical surface with infinite length.
In 1N and 1B phases, the interface is planar surface with infinite area.
The unit cells have a finite lattice period along three spatial dimensions (for 3N and 3B), or along two dimensions (for 2N and 2B), or along only a single dimension (for 1N).
Notice that the variables $u=V_N/V_c$, Eq. (\ref{def_u}), and $u^{\rm bub}$, Eq. (\ref{def_u_BUB}) (which is equal to the ratio between volumes of the bubble and the unit cell), are essentially different.

As the parameters relevant to the structure have been defined, let us turn to the energy of the system.
The basic quantities in the theory actually are the energy densities averaged on length scales much larger than the volume of the unit cell $V_c$.
In this paper, the energy densities are denoted by $w$, while the energies by $E$.
The tilde is used above any quantity which is a function of space, in contrast to quantities averaged over the unit cell which have no tilde.

\subsection{Total energy in the unit cell}
The total energy and the total energy density are related in the unit cell as following:
\begin{equation}\label{deltaEtot}
   E_{\mathrm{tot}}= \int\mathbf{d^3r}\;\tilde{w}_{\mathrm{tot}}[r_N,\tilde{n}(\mathbf{r}),\tilde{x}(\mathbf{r}),\tilde{n}_{no}(\mathbf{r})].
\end{equation}
Let us average the functions over the unit cell, which allows us to consider a set of parameters rather than the functions, therefore, we can work with the averaged total energy:
\begin{equation}\label{deltaEtot}
   E_{\mathrm{tot}}= V_c{w}_{\mathrm{tot}}(r_N,{x},{n}_{ni},u).
\end{equation}
The total energy density for 1N, 2N and 3N phases is a sum of the binding, the Coulomb, the pure neutron liquid and the electron contributions:
\begin{equation}\label{wtotSect1}
w_{\mathrm{tot}}=w_{\mathrm{B}} + w_{\mathrm{C+L}} + w_{\mathrm{no}} + w_{\mathrm{e}}.
\end{equation}

The binding contribution consists of the bulk (uniform) energy density of the nucleus and of the part associated with the surface of the nucleus:
\begin{equation}\label{wB}
w_{\mathrm{B}}=w_{\mathrm{nuc}} + w_{\mathrm{s}} + w_{\mathrm{cur}}.
\end{equation}
The bulk energy density of the nucleus is
\begin{equation}\label{def_wnuc}
  w_{\rm nuc}=un\left[\left(1-{x}\right)m_n + {x}m_p\right]c^2 + u{n}\varepsilon({n},{x}),
\end{equation}
where $m_p$ and $m_n$ are the proton and neutron rest masses, correspondingly, $c$ is the speed of light and $\varepsilon$ is equal to the energy per baryon in uniform nuclear matter inside the nucleus and zero outside.

The plain surface energy density $w_{\rm s}$ is proportional to the surface area of the nucleus and therefore has the form
\begin{equation}\label{def_ws}
  w_{\rm s}=\frac{ud}{r_N}\sigma_s(x),
\end{equation}
where $\sigma_s(x)$ is a function with dimension MeV fm$^{-2}$.

The energy density associated with curvature of the nucleus surface $w_{\mathrm{cur}}$ is proportional to both the surface area and to the mean principal curvature, which is $2/r_N$ for nuclei in 3N phase, $1/r_N$ for those in 2N phase and 0 in 1N phase:
\begin{equation}\label{def_wbend}
  w_{\rm cur}=\frac{ud(d-1)}{r_N^2}\sigma_c,
\end{equation}
where $\sigma_c=\sigma_c(x)$ is a parameter with dimension MeV fm$^{-1}$.

The Coulomb energy density including the nucleus self-energy and the lattice energy was found in \cite{RavenhallPethickWilson1983,PethickRavenhall1995} and is given in
\begin{equation}\label{def_wCoul}
  w_{\rm C+L}=2\pi(enxr_N)^2uf_d(u),
\end{equation}
where
\begin{equation}\label{def_fdu}
  f_d(u)=\frac{1}{d+2}\left[\frac{2}{d-2}\left(1-\frac{du^{1-2/d}}{2}\right) + u\right],
\end{equation}
with $f_3(u)=(1/5)(2-3u^{1/3}+u)$, $f_2(u)=(1/4)\left(-\ln{u}-1+u\right)$ and $f_1(u)=(1/3)\left(1/u-2+u\right)$.
The energy density of the dripped neutrons $w_{\mathrm{no}}$ is
\begin{equation}\label{def_wno}
  w_{\mathrm{no}}=\left(1-u\right){n}_{no}\left[m_nc^2 + \varepsilon({n}_{no},0)\right].
\end{equation}
The electron energy density $w_{\rm e}$ is
\begin{equation}\label{def_we}
  w_{\rm e}=\frac{3}{4}\hbar c(3\pi^2)^{1/3} (u{n}{x})^{4/3},
\end{equation}
where he electrical neutrality condition in the unit cell has been used, so the electron number density $n_e$ averaged on $V_c$ is equal to $n_e=unx$.
\subsection{Basic equations}
To derive the basic equations that determine the thermodynamic equilibrium state of the system, the variational calculus is applied to minimize the functional $E_{\mathrm{tot}}$:
\begin{equation}\label{deltaEtot}
  \delta E_{\mathrm{tot}} = 0.
\end{equation}
The reason to use the set defined in Eq. (\ref{VarSet2}) rather than the one defined in Eq. (\ref{VarSet1}), is that the resulting variational equations have clear physical significance, as has been revealed in \cite{WatanabeEtAl2000}.

The variational equations have the form
\begin{equation}\label{def_VarEq}
  \frac{\delta w_{\rm {tot}}}{\delta Y}=0,
\end{equation}
where $Y$ is either of the parameters from the set defined in Eq. (\ref{VarSet2}).
As explained in Appendix, from  Eq. (\ref{def_VarEq}) one obtains four basic equations of the CLDM:
\begin{eqnarray}
  && \label{var_1} w_{\rm s} + 2w_{\rm cur}=2w_{\rm C+L}, \\
  && \label{var_2} \mu_{e}=\mu_{ni} - \mu_{pi}, \\
  && \label{var_3} \mu_{ni}=\mu_{no}, \\
  && \label{var_4} P_i=P_o.
\end{eqnarray}
The explicit form of the quantities in Eqs. (\ref{var_2})-(\ref{var_4}) is given in Eqs. (\ref{def_mu_cluster}), (\ref{Mupi}), (\ref{Muni}), (\ref{Muno}), (\ref{Pi}), (\ref{Po}).

The first of these equations, Eq. (\ref{var_1}), is sometimes called the nuclear energy law and corresponds to $Y=r_N$ in Eq. (\ref{def_VarEq}).
Equation (\ref{var_2}) is the condition of the chemical equilibrium and corresponds to $Y={x}$ in Eq. (\ref{def_VarEq}).
The third equation, Eq. (\ref{var_3}), states that the energy to add a neutron to the nucleus is equal to the energy to add a neutron to the pure neutron liquid and results from Eq. (\ref{def_VarEq}) with $Y={n}_{ni}$.
The fourth equation, Eq. (\ref{var_4}), follows from Eq. (\ref{def_VarEq}) with $Y=u$ and implies the equality of pressures inside and outside the nucleus.
Notice that pressure inside the nucleus includes the surface and Coulomb contributions.

\section{Graphical method of solution of the CLDM equations}
Solution of the basic equations of the CLDM, Eqs. (\ref{var_1})-(\ref{var_4}) can be achieved in few steps.
The first is to express $r_N$ from Eq. (\ref{var_1}) analytically for given $n_b$, $n_{no}$, $n$ and $x$.
The second step is to find the equilibrium proton fraction $x$ from Eq. (\ref{var_2}), which requires a numerical approximation.
In this step, a range of relevant values of the baryon number density inside the nucleus $n$ is chosen empirically, for instance, $0.5<\eta<1.5$, where
\begin{equation}\label{def_eta}
  \eta\equiv\frac{n}{n_0}.
\end{equation}
The numerical values for $n_{no}$ are also chosen empirically.

To the best of my knowledge, the standard procedure to solve Eq. (\ref{var_1}) has been to start from the analytical solution of equation without the curvature, $w_{\rm s} = 2w_{\rm C+L}$ and then to find the solution including the curvature numerically and assuming it is small, see Eqs. (6)-(12) in \cite{PethickRavenhallLattimer1983}.
A similar approximate scheme of solution has been used recently in Eqs. (15)-(18) of \cite{NakazatoIidaOyamatsu2011}.
However, the curvature correction $w_{\rm cur}$ is not very small.
More generally, an analytical solution, if known, is preferable over the corresponding numerical approximation.

In this paper, the theory is advanced by using the complete analytical solution to Eq. (\ref{var_1}) valid for any values of $w_{\rm cur}$.
To obtain this solution, I insert the explicit form of the terms in Eq. (\ref{var_1}) and find the equation that determines $r_N$:
\begin{equation}\label{rNequation}
  r_N^4 - 4qr_N - 3r = 0,
\end{equation}
where
\begin{eqnarray}
\label{qpoly} q = \frac{1}{4}\frac{\sigma_s(x)d}{4\pi(enx)^2f_d(u)}, \\
\label{rpoly} r = \frac{1}{3}\frac{2d(d-1)\sigma_c(x)}{4\pi(enx)^2f_d(u)}.
\end{eqnarray}
As it is explained in Appendix E, the solution to Eq. (\ref{rNequation}) is given by the fourth-order polynomial root:
\begin{equation}\label{solution_rN}
  r_N=\sqrt{\frac{p}{2}}\left[1 + \sqrt{q\left(\frac{2}{p}\right)^{3/2}-1}\right],
\end{equation}
where
\begin{equation}\label{p}
  p=\left[q^2 + \sqrt{q^4+r^3}\right]^{1/3} - \left[\sqrt{q^4+r^3}-q^2\right]^{1/3}.
\end{equation}
In case when the curvature contribution is neglected, Eq. (\ref{solution_rN}) reduces to
\begin{equation}\label{solution_rN_noCurvature}
  r_N=\left[\frac{\sigma_s(x) d}{4\pi(enx)^2f_d(u)}\right]^{1/3}, \quad(\sigma_c=0).
\end{equation}

In order to solve Eqs. (\ref{var_2})-(\ref{var_4}), I first express the volume fraction $u$ from Eqs. (\ref{def_nb}) as
\begin{equation}\label{u_expr}
  u=\frac{n_b-n_{no}}{n-n_{no}},
\end{equation}
which allows to find $r_N$ using Eqs. (\ref{solution_rN}).

For each numerical pair of $(n,n_{no})$ I find $x(n,n_{no})$, which satisfies Eq. (\ref{var_2}).
In the next step I find, for each value of $n$ the value of $n_{no}(n)$ from Eqs. (\ref{var_3}) and (\ref{var_4}), resulting in two distinct curves, $n_{no}=n_{no}^{P}(n)$ and $n_{no}=n_{no}^{\mu}(n)$.
For values corresponding to the curve $n_{no}=n_{no}^{P}(n)$, Eqs. (\ref{var_1}), (\ref{var_2}) and (\ref{var_3}) are satisfied.
Conversely, for values corresponding to the curve $n_{no}=n_{no}^{\mu}(n)$, Eqs. (\ref{var_1}), (\ref{var_2}) and (\ref{var_4}) are satisfied.
In the final step, I search for the intersection of these two curves.
If the intersection point exists, then the full set of the CLDM Eqs. (\ref{var_1})-(\ref{var_4}) is satisfied at the intersection point.

\begin{widetext}
\section{Numerical evaluation of basic parameters}
\subsection{The uniform nuclear energy}
To compute the uniform contribution $\varepsilon$ in Eq. (\ref{def_EB_I}), the Skyrme mean-field model based on the chiral effective field theory and constrained by the ground-state energies of doubly magic nuclei and observations of neutron stars \cite{LimHolt2017} is used.
The numerical data was kindly provided by Xavier Vi\~nas \cite{KobyakovVinas2024}.
The mean field is represented by the nucleon densities $\tilde{n}_q=\tilde{n}_q(\mathbf{r})$ ($q=n,p$), which are real-valued functions of the spatial coordinates.
This approach is also applicable to nonuniform cases and will be used below for evaluation of the surface properties.

The quantum kinetic energy density is replaced by the semiclassical counterpart up to terms of the order $\hbar^2$ resulting second order derivatives of the neutron and proton densities \cite{BrackGuetHakanson1985}.
The energy density reads:
\begin{eqnarray}
&& \nonumber \tilde{w} = \frac{\hbar^2}{2m}(f_n\tau_n+ f_p\tau_p) + \frac{t_0}{4}[(x_0+2)\tilde{n}^2 -(2x_0+1)(\tilde{n}_n^2+\tilde{n}_p^2)]-\frac{W_0}{2} (\tilde{n} {\bf \nabla}\cdot{\bf J} + \tilde{n}_n {\bf \nabla}\cdot{\bf J_n} + \tilde{n}_p {\bf \nabla}\cdot{\bf J_p}) \\
&& \nonumber - \frac{1}{32}[t_2(2+x_2)-3t_1(2+x_1)](\nabla \tilde{n})^2  - \frac{1}{32}[3t_1(2x_1+1)+t_2(2x_2+1)]
[(\nabla \tilde{n}_n)^2+(\nabla \tilde{n}_p)^2] \\
&& \label{eqA11} +\frac{t_3 \tilde{n}^{\alpha_1}}{24}[(x_3+2)\tilde{n}^2 -(2x_3+1)(\tilde{n}_n^2+\tilde{n}_p^2)]
+ \frac{t_4 \tilde{n}^{\alpha_2}}{24}[(x_4+2)\tilde{n}^2 -(2x_4+1)(\tilde{n}_n^2+\tilde{n}_p^2)],
\end{eqnarray}
where $\tilde{n}=\tilde{n}_p+\tilde{n}_n$. The scaled inverse of the nucleon effective mass for each nucleon species $f_q$ is defined as
\begin{equation}
f_q = \frac{m}{\tilde{m_q}} = 1 + \frac{m}{4\hbar^2}\{[t_1(x_1+2) + t_2(x_2+2)]\tilde{n}
+ [t_2(2x_2+1) - t_1(2x_1+1)]\tilde{n}_q \}.
\label{eqA12}
\end{equation}
Here, $\tau_q$ and ${\bf J}_q$ are the $\hbar^2-$order kinetic energy density and the spin densities, respectively, which read
\begin{eqnarray}
&& \tau_q  =  \frac{3}{5}(3\pi^2)^{2/3} \tilde{n}_q^{5/3} + \frac{1}{36}\frac{({\bf \nabla}\tilde{n}_q)^2}{\tilde{n}_q}
- \frac{1}{3} \frac{{\bf \nabla}f_q{\bf \nabla}\tilde{n}_q}{f_q}  - \frac{1}{12} \tilde{n}_q\frac{({\bf \nabla}f_q)^2}{f^2_q}
+ \frac{({\bf J}_q)^2}{2 \tilde{n}_q}, \\
&& {\bf J}_q = \frac{m}{\hbar^2}W_0 \frac{n_q}{f_q} ({\bf \nabla}n + {\bf \nabla}n_q)
\label{eqA13}
\end{eqnarray}
Mapping of the spatial-dependent energy $\tilde{w}$ in Eq. (\ref{eqA11}) to the energy per baryon $\varepsilon$ in uniform nuclear matter can be done using Eq. (1) of \cite{Dutra2012} yielding
\begin{eqnarray}
 \nonumber && \varepsilon(n,x)=\frac{3}{5}\varepsilon_0(2\eta)^{\frac{2}{3}}\left\{  \left[x^{\frac{5}{3}}+(1-x)^{\frac{5}{3}}\right](1+\frac{\tilde{a}}{8}\eta) + \left[x^{\frac{8}{3}}+(1-x)^{\frac{8}{3}}\right]\frac{\tilde{b}}{4}\eta  \right\}         \\
 \nonumber && +\frac{t_0n_0}{4}\eta  \left\{ x_0+2-(2x_0+1)\left[x^2+(1-x)^2\right] \right\}\\
 \nonumber && +\frac{t_3n_0^{\alpha_1+1}}{24}\eta^{\alpha_1+1}  \left\{ x_3+2-(2x_3+1)\left[x^2+(1-x)^2\right] \right\}\\
 \label{def_exn} && +\frac{t_4n_0^{\alpha_2+1}}{24}\eta^{\alpha_2+1}  \left\{ x_4+2-(2x_4+1)\left[x^2+(1-x)^2\right] \right\}  ,
\end{eqnarray}
where $\eta$ is given in Eq. (\ref{def_eta}) and
\begin{equation}
 \label{def_a} \tilde{a}=\frac{2m_p}{\hbar^2}n_0\left[t_1(x_1+2)+t_2(x_2+2)\right],\quad \tilde{b}=\frac{2m_p}{\hbar^2}n_0\left[t_2(2x_2+1)-t_1(2x_1+1)\right].
\end{equation}
\end{widetext}
The numerical values of the parameters are given in Table II.
The critical density at which the uniform nuclear matter becomes unstable with respect to the hydrostatic perturbations is
\begin{equation}\label{def_eta_uni_star}
\eta_{\rm uni}^{*}\equiv n_t/n_0,
\end{equation}
with $n_t$ found in Eq. (30) of \cite{LimHolt2017}.

The nuclear saturation condition expresses the fact that the symmetric nuclear matter at saturation is pressureless:
\begin{equation}\label{PnucSaturation}
P_{\rm nuc}(n=n_0,x=1/2)=0.
\end{equation}
The nuclear binding energy per nucleon reads
\begin{equation}\label{def_B}
B=\varepsilon(n=n_0,x=1/2).
\end{equation}
The nuclear symmetry energy is defined as
\begin{equation}\label{def_S}
S=\varepsilon(n=n_0,x=0)-\varepsilon(n=n_0,x=\frac{1}{2})
\end{equation}
and its slope parameter
\begin{equation}\label{def_L}
L=\frac{3}{8}n_0\partial^3_{nxx}\varepsilon|_{n=n_0,\,x=1/2}.
\end{equation}
The incompressibility reads
\begin{equation}\label{def_K}
K=9\partial_{n}P_{\rm nuc}|_{n=n_0,\,x=1/2}.
\end{equation}
The parameters $K$ and $S$ have been constrained by the experiment: the expected range of $K$ is $230\pm30$ MeV and the expected range of $S$ is $32\pm2$ MeV.
However, the slope parameter $L$ is not well constrained as various models predict different values of $L$.
For instance, the Skyrme interactions favored in \cite{Dutra2012} predict $L$ in the range between 53.04 and and 61.45 MeV; the chiral effective field theory predictions reported in \cite{GramsEtal2022} span the range between 36.5 and 69.0 MeV.
\subsection{Surface and curvature tensions}
Surface tension of the nucleus is the excess total energy per area associated with non-uniformity of the density profile of nuclear matter at the edge of the nucleus.
As a part of the total energy, the potential energy can be expressed in terms of the nucleon densities in uniform nuclear matter $n_n$ and $n_p$, and the square of their gradients, $({\bf \nabla}n_n)^2$ and $({\bf \nabla}n_p)^2$, which simulate the finite range of the nucleon-nucleon interaction.
The remaining part of the total energy, the kinetic energy, is expanded in powers of $\hbar^2$.
We take only the two first terms of this expansion.
The leading order term corresponds to the bulk, which includes the Thomas-Fermi kinetic energy densities for neutrons and protons.
The next-to-leading order term contains the second order gradients coming from the Skyrme force and the Weizs$\ddot{\mathrm{a}}$cker correction to the kinetic energy density and effective masses, as well as a contribution of the semiclassical spin density.

In the lowest approximation the nucleus is simulated as a half-space filled with uniform nuclear matter \cite{RavenhallBennettPethick1972,RavenhallPethickLattimer1983}.
In this subsection, we choose the new coordinates, where the nuclear matter is located at $z<0$ the pure neutron liquid at $z>0$.
The surface tension is defined as \cite{CentelesEstalVinas1998}
\begin{equation}
\sigma_s = \int^{+ \infty}_{- \infty} dz\;\left[ P_0 - P(z) \right],
\label{SurTen1}
\end{equation}
where $\mu_n$ and $\mu_p$ are the chemical potentials, $\tilde{w}(z)$ is the local energy density of the system defined in Eq. (\ref{eqA11}), $w_{no}=\tilde{w}(z \to + \infty)$ is the energy density of the pure neutron liquid and $P(z)= \mu_n \tilde{n}_n(z) + \mu_p \tilde{n}_p (z) - \tilde{w} (z)$, $P_0=\mu_n n_{no} - w_{no}$.
The limit $\tilde{\varepsilon} (z \to - \infty)$,  corresponds to the energy density of uniform asymmetric nuclear matter.
The number density of neutrons is $\tilde{n}_n(z)$, that of protons is $\tilde{n}_p(z)$ and the number density of neutrons in the bulk of the pure neutron liquid ${n}_{no}=\tilde{n}_n(z \to + \infty)$.
The functions $\tilde{n}_n(z)$ and $\tilde{n}_p(z)$ are the solutions of the Euler-Lagrange equations derived from Eq. (\ref{SurTen1}):
\begin{equation}
\frac{\delta \tilde{w}}{\delta \tilde{n}_n} - \mu_n =0, \quad  \frac{\delta \tilde{w}}{\delta \tilde{n}_p} - \mu_p =0, \quad
\frac{\delta {w}_{no}}{\delta n_{no}} - \mu_n =0.
\label{SurTen2}
\end{equation}

The curvature tension is a correction linear in the principal curvature, which reads \cite{CentelesEstalVinas1998,DouchinHaenselMeyer2000}:
\begin{equation}
\sigma_c = \int^{+ \infty}_{- \infty} dz (z-z_p)\left[ P_0 - P(z) \right],
\label{eqA7}
\end{equation}
where $z_p$ is the proton surface location, which is defined as
\begin{equation}\label{def_zp}
\int^{+ \infty}_{- \infty} dz (z - z_p) \frac{d \tilde{n}_p(z)}{dz} = 0.
\end{equation}
Equations (\ref{SurTen2}) are solved using the imaginary timestep method.
The solutions enable us to compute the surface $\sigma_s$ and the curvature $\sigma_c$ tensions from Eqs. (\ref{SurTen1}) and (\ref{eqA7}).

\section{Numerical results}
\subsection{Calculation of $\sigma_s$ and $\sigma_c$}
Here I present novel results on the surface $\sigma_s$ and the curvature $\sigma_c$ tensions along with their derivatives with respect to $x$, $\sigma_s'$ and $\sigma_c'$, computed numerically in the Thomas-Fermi approximation from the Skyrme mean-field model based on the chiral effective field theory.
The reason that the derivatives are relevant can be seen from the explicit form of the basic Eqs. (\ref{var_2}) and (\ref{var_3}) given in Eqs. (\ref{Muni}) and (\ref{Mupi}).

Figures 1 and 2 show the numerical results.
For convenience, the data on the surface tensions is also presented in Table III.
To the best of my knowledge, the numerical results shown in Table III are novel.
\begin{figure}
\includegraphics[width=3.5in]{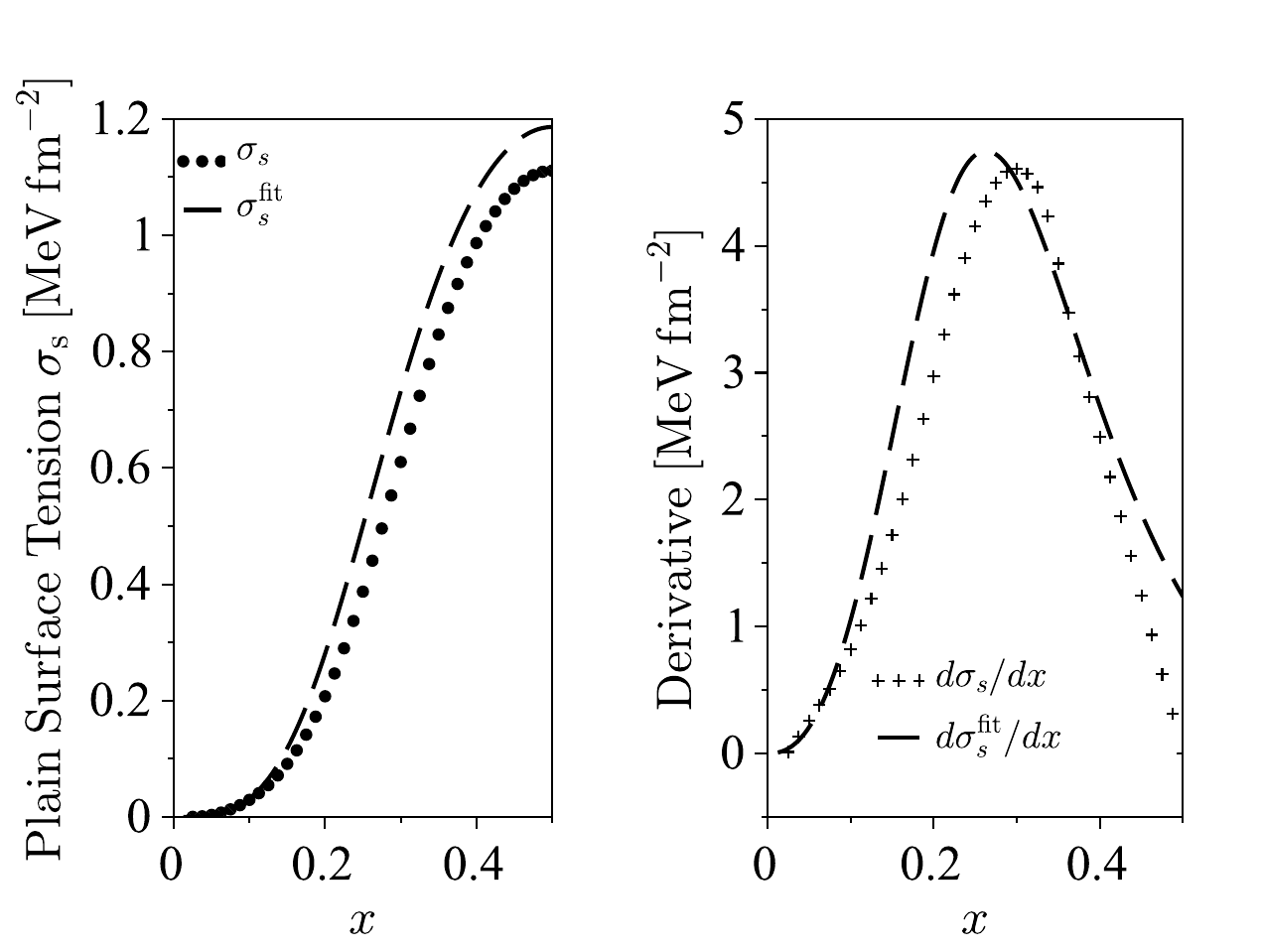}
\caption{The plain surface tension $\sigma_s$ and its derivative $d\sigma_s/dx$ computed from Eq. (\ref{SurTen1}) and the numerical fits $\sigma_s^{\rm fit}$ and $d\sigma_s^{\rm fit}/dx$ according to \cite{LimHolt2017}. Markers show the data from Table III. Lines show Eq. (\ref{def_sigmaLimHolt}) and its derivative.}
\end{figure}
\begin{figure}
\includegraphics[width=3.5in]{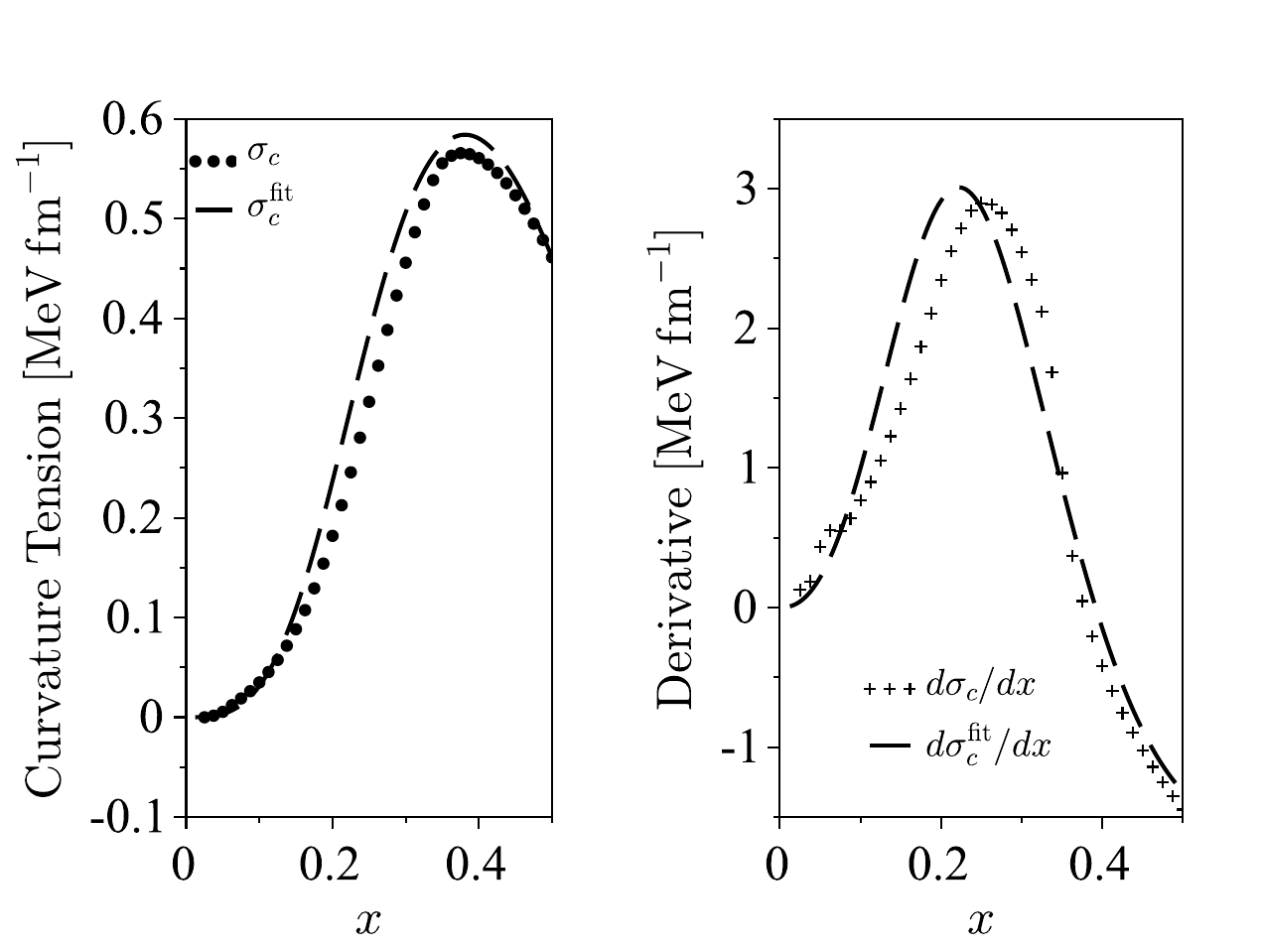}
\caption{The curvature tension $\sigma_c$ and its derivative $d\sigma_c/dx$ computed from Eq. (\ref{eqA7}) and the numerical fits $\sigma_c^{\rm fit}$ and $d\sigma_c^{\rm fit}/dx$ according to Eq. (\ref{def_sigmaCfit}). Markers show the data from Table III. Lines show Eq. (\ref{def_sigmaCfit}) and its derivative.}
\end{figure}

\subsection{Actual solution of the basic equations}
Using the method described above, I solve Eqs. (\ref{var_1})-(\ref{var_4})) and display in Figs. 3-6 the solutions corresponding the complete set of quantities defined in Eq. (\ref{VarSet1}).
The basic equations are solved for each of the possible pasta phases discussed above, whenever the solution exists.
\begin{figure}
\includegraphics[width=3.5in]{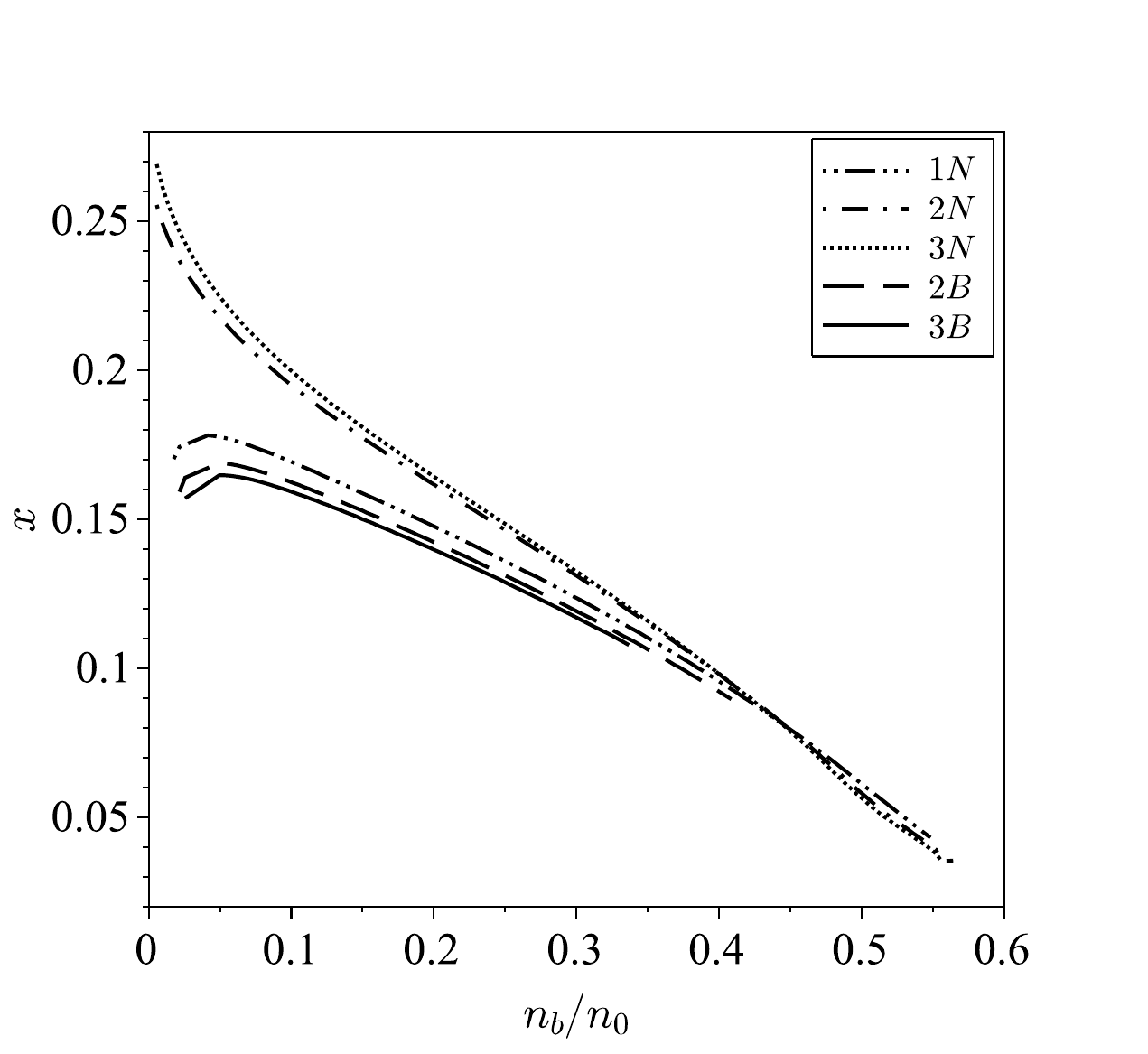}
\caption{Solution to the CLDM equilibrium, Eqs. (\ref{var_1})-(\ref{var_4}): The proton fraction $x$ inside the nucleus,
Eq. (\ref{def_x}), and outside the bubble, Eq. (\ref{def_x_BUB}), as function of the mean baryon density $n_b$ scaled by $n_0$.}
\end{figure}
\begin{figure}
\includegraphics[width=3.5in]{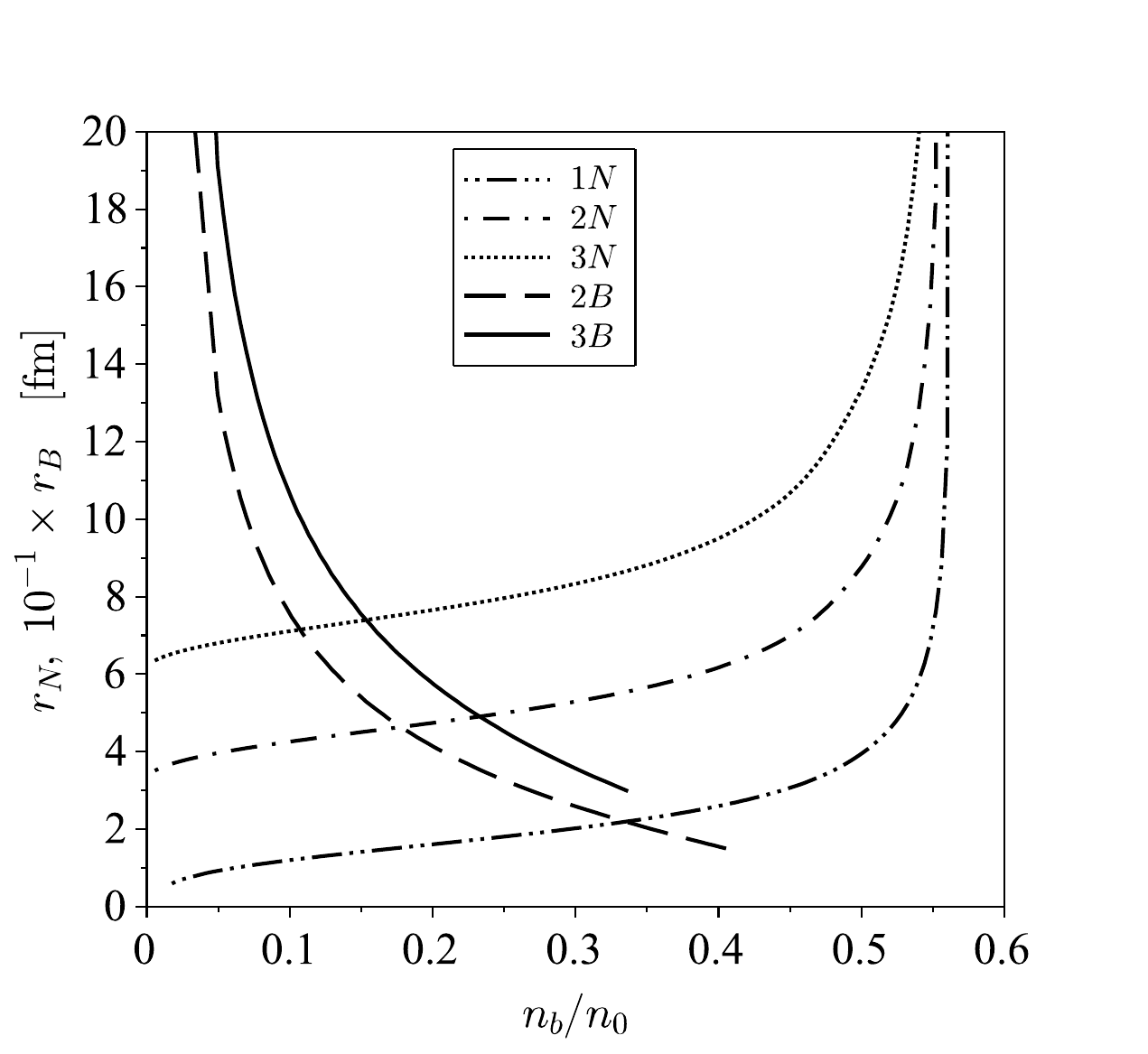}
\caption{Solution to the CLDM equilibrium, Eqs. (\ref{var_1})-(\ref{var_4}): The nucleus radius $r_N$ and the bubble radius
$r_B$ scaled by a factor of 10).}
\end{figure}
\begin{figure}
\includegraphics[width=3.5in]{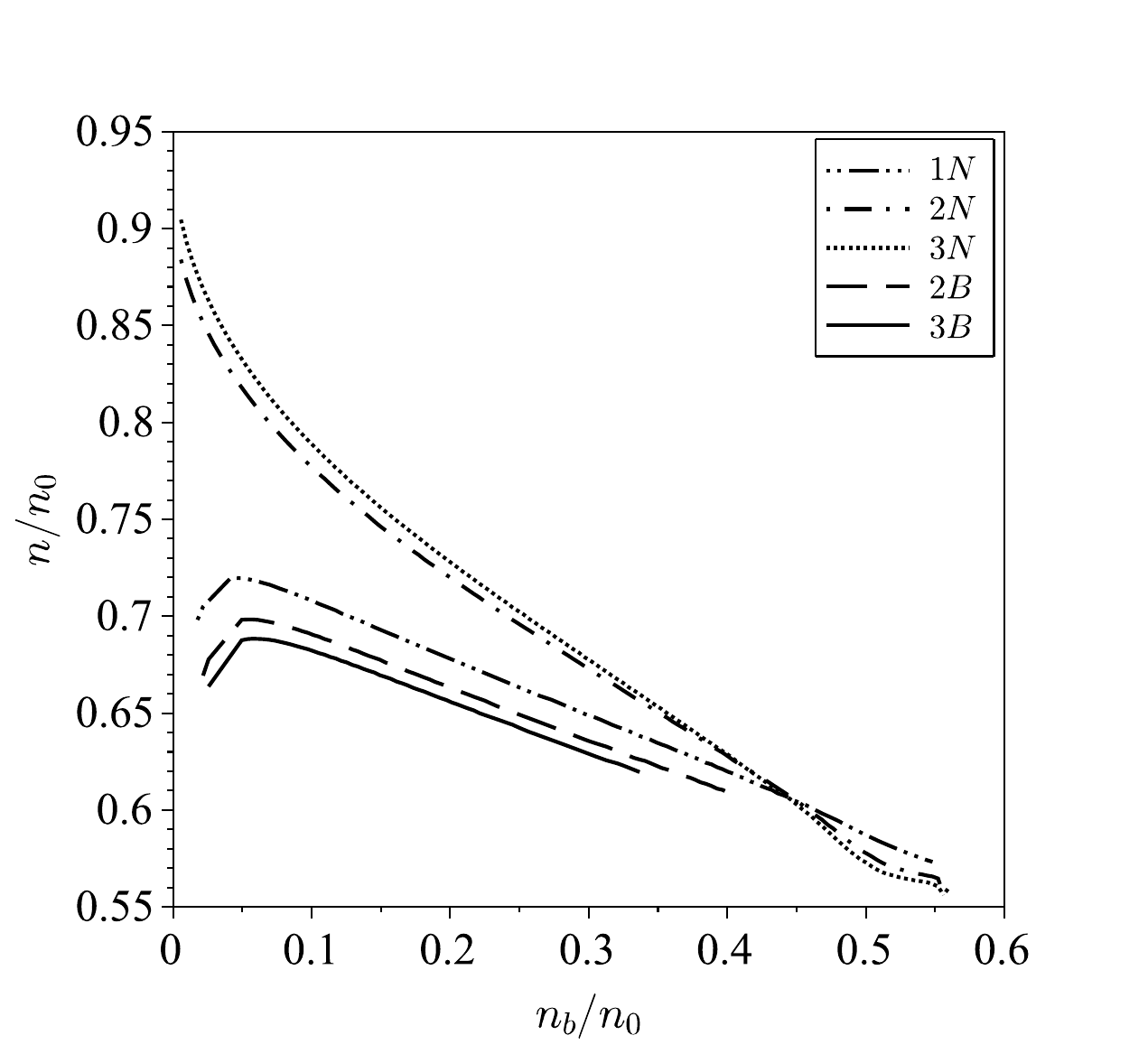}
\caption{Solution to the CLDM equilibrium, Eqs. (\ref{var_1})-(\ref{var_4}): The baryon number density inside the nucleus and outside
 the bubble.}
\end{figure}
\begin{figure}
\includegraphics[width=3.5in]{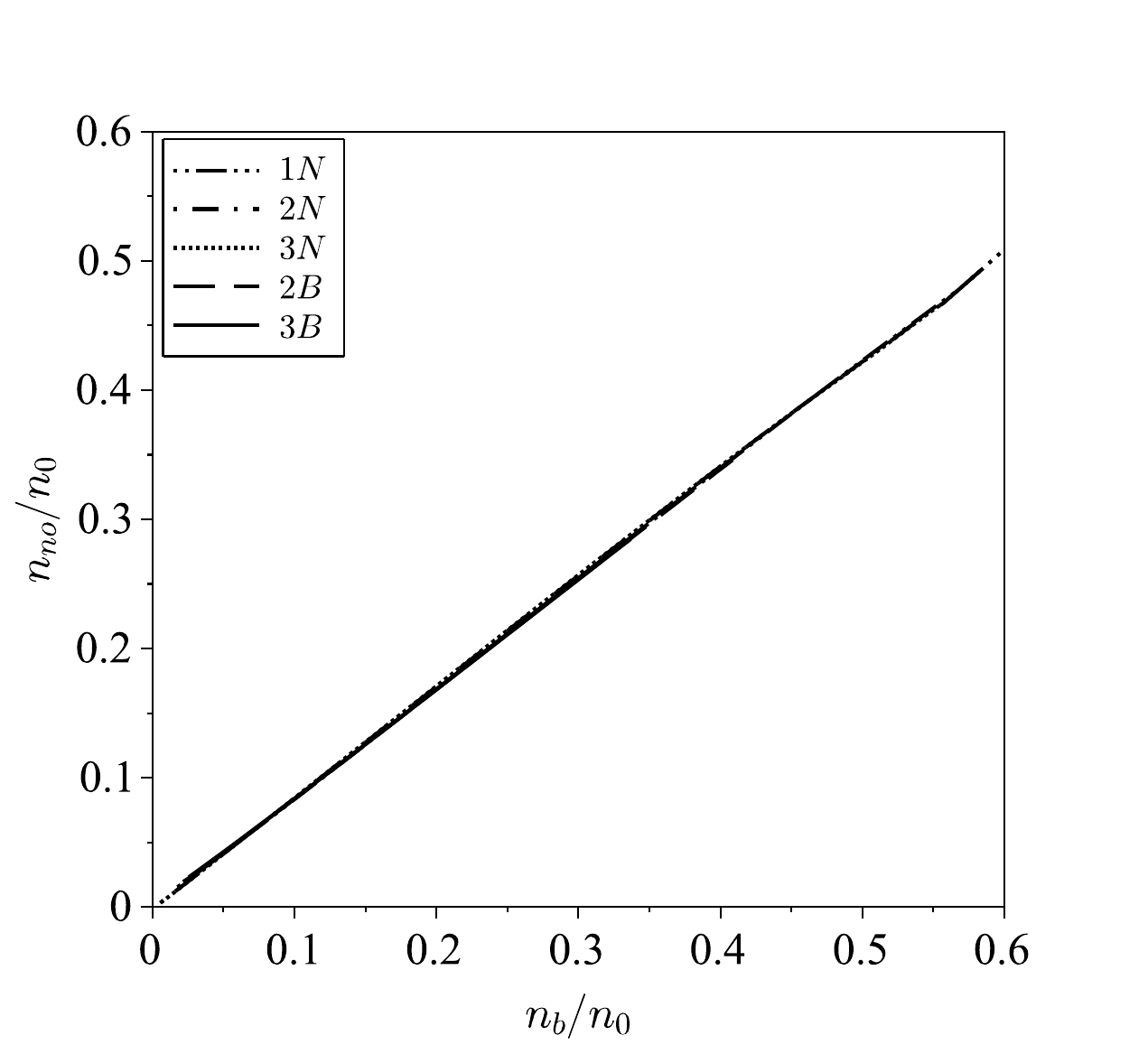}
\caption{Solution to the CLDM equilibrium, Eqs. (\ref{var_1})-(\ref{var_4}): The baryon number density $n_{no}$ of pure neutron
matter outside the nucleus and inside the bubble.}
\end{figure}

In order to check independently the numerical results, I calculate $w_{\rm tot}$ for each pasta phase and find the phase $\tau$ corresponding to the smallest energy density as function of the mean baryon number density $n_b$, where $\tau=1N$, $2N$, $3N$, $2B$, $3B$ or $\rm{Uni}$ denotes lasagna, spaghetti, spherical nuclei, rod-like bubbles, spherical bubbles or uniform nuclear matter.
Assuming that the ground state of the pasta phase is perfectly ordered, the function $\tau(n_b)$ represents the ground state of the inner crust.
The black line in Fig. 7 shows the result of calculation of $\tau(n_b)$ from the CLDM with the bulk part $\varepsilon$ obtained from the same nucleon force parametrization as used in \cite{LimHolt2017} and with the surface properties $\sigma_s$ and $\sigma_c$ obtained here self-consistently as shows Table III (this is Case 4 in Table I).

In Fig. 7, the difference between the present predictions for the ground state and those suggested in \cite{LimHolt2017} is qualitative.
This is explained as a direct effect of the inclusion of $\sigma_c$ in the present calculations.
\begin{figure}
\includegraphics[width=3.5in]{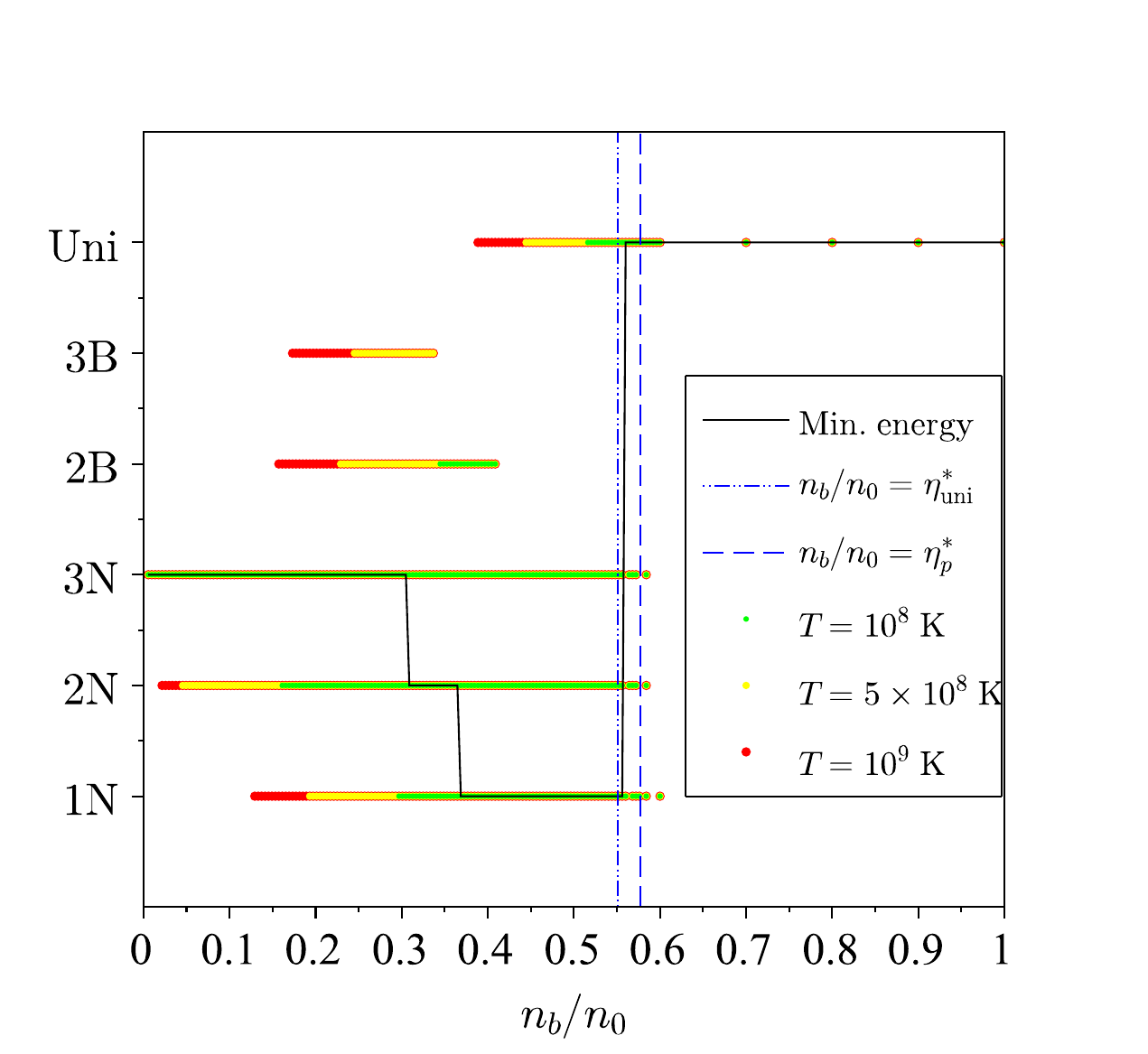}
\caption{(Color online) The ground state of the inner crust (black solid line) and quasi-equilibrium ordered pasta configurations with higher energy per baryon separated from the ground state by less than $k_BT$ (color dots, red for $T=10^9$ K, gold for $T=5\times10^8$ K and green for $T=10^8$ K).
The transitions between various pasta phases are seen as vertical jumps of the black solid line.
The dash-dotted line marks the critical value of $n_b$ at which uniform nuclear matter becomes unstable with respect to long wavelength
hydrodynamic fluctuations (see Eq. (\ref{def_eta_uni_star}) for details).
The dashed line marks the critical value of $n_b$ at which protons start dripping out from the nuclei (see Fig. 8 for details).}
\end{figure}

Table I demonstrates explicitly the influence of the curvature contribution, where the results for the phase transition densities between different pasta phases are presented.
Case 1 shows the results obtained in \cite{LimHolt2017}.
Case 2 presents the results obtained in our numerical algorithm when I use the same input parameters which were used in \cite{LimHolt2017}, specifically, I set $\sigma_c=0$ and for $\sigma_s$ use the trial fitting function given in Eqs. (\ref{def_sigmaLimHolt}) and (\ref{parametersLimHolt}).
Confronting the outputs from Cases 1 and 2, one observes that the present calculations independently reproduce the results of \cite{LimHolt2017} with a good precision.

In order to understand the sensitivity of the transition densities on the values of $\sigma_s$ alone, I have solved the CLDM equations replacing the fit (Eqs. (\ref{def_sigmaLimHolt}) and (\ref{parametersLimHolt})) for $\sigma_s$ by the self-consistent data (Table III) for $\sigma_s$, while keeping zero curvature tension $\sigma_c=0$ (this is Case 3 in Table I).

Case 4 in devoted to the main calculations of this paper, when the CLDM equations are solved with the input data for the surface tension and the curvature tension provided in Table III.
\begin{table}
\begin{tabular}{|c|c|c|c|c|}
  \hline
                       & $\eta$ (Case 1) & $\eta$ (Case 2)  & $\eta$ (Case 3)   & $\eta$ (Case 4) \\
  \hline
  3N-2N                & 0.4061          & 0.4025$\pm$0.002 & 0.4145$\pm$0.002  & 0.3068$\pm$0.002 \\
  \hline
  2N-1N                & 0.4715          & 0.4664$\pm$0.002 & 0.4704$\pm$0.002  & 0.3666$\pm$0.002 \\
  \hline
  1N-2B                & 0.5361          & 0.5342$\pm$0.002 & 0.5302$\pm$0.002  & -- \\
  \hline
  2B-3B                & 0.5502          & 0.5501$\pm$0.002 & 0.5462$\pm$0.002  & -- \\
  \hline
  3B-Uni.              & 0.5624          & 0.5621$\pm$0.002 & 0.5661$\pm$0.002  & -- \\
  \hline
  1N-Uni.              & --              & --               &  --               &  0.5581$\pm$0.002 \\
  \hline
  $\eta_{\rm uni}^{*}$ & --              & 0.5508           &  0.5508           & 0.5508 \\
  \hline
  $\eta_{\rm p}^{*}$   & --              & 0.5661$\pm$0.002 &  0.5741$\pm$0.002 &  0.5741$\pm$0.002 \\
  \hline
\end{tabular}
\caption{The baryon number densities $\eta\equiv n_b/n_0$ at which the solution with the minimum energy per baryon changes its phase between 1N, 2N, 3N, 2B, 3B and the uniform nuclear matter.
The scaled baryon densities $\eta_{\rm uni}^{*}$ and $\eta_{\rm p}^{*}$ indicate when the uniform nuclear matter is unstable with respect to
modulations of the nucleon densities and when the proton drip occurs.
Case 1 is the direct output data from Table III of \cite{LimHolt2017}.
Case 2 shows computationally reproduced results of \cite{LimHolt2017} obtained by solving to Eqs. (\ref{var_1})-(\ref{var_4}) with $\sigma_c=0$ and with $\sigma_s$ from Eqs. (\ref{def_sigmaLimHolt}) and the fitting parameters from Eq. (\ref{parametersLimHolt}).
Case 3 is the solution to Eqs. (\ref{var_1})-(\ref{var_4}) with $\sigma_c=0$ and with $\sigma_s$ from Table III.
Case 4 is the complete solution to Eqs. (\ref{var_1})-(\ref{var_4}) with $\sigma_s$ and $\sigma_c$ from Table III.
}
\end{table}

Two bottom lines in Table I display the high-density properties of the CLDM related to the transition from nonuniform to the uniform nuclear matter.
This phase transition can be found by considering the hydrostatic stability of the uniform nuclear matter as the baryon density is decreased.
The corresponding value of the (scaled) mean baryon density is denoted as $\eta_{\rm uni}^{*}$, which is computed using Eq. (30) and Table V of \cite{LimHolt2017}.
From the other hand, the CLDM predicts the value of the proton chemical potential inside the nucleus $\mu_{pi}$ and outside $\mu_{po}$ as functions of $n_b$.
The uniform nuclear matter is thermodynamically favored over the pasta phases when $\Delta\mu_p\leq0$ is satisfied, where
\begin{equation}\label{def_deltaMuP}
      \Delta\mu_p=\mu_{po}-\mu_{pi}.
\end{equation}
In order to find $\eta_p^{*}$, where $n_0\eta_p^{*}$ is the mean baryon density when $\Delta\mu_p$ becomes nonpositive, I calculate this quantity numerically from the solutions of the CLDM shown in Figs. 3-6.
$\Delta\mu_p$ represents the workfunction of protons, which is a potential energy barrier created by the cluster, that prevents protons from flowing out of the nuclear cluster into the pure neutron matter.

Figure 8 shows $\Delta\mu_p$ as function of the mean baryon density $n_b$, for five possible pasta phases in the inner crust.
\begin{figure}
\includegraphics[width=3.5in]{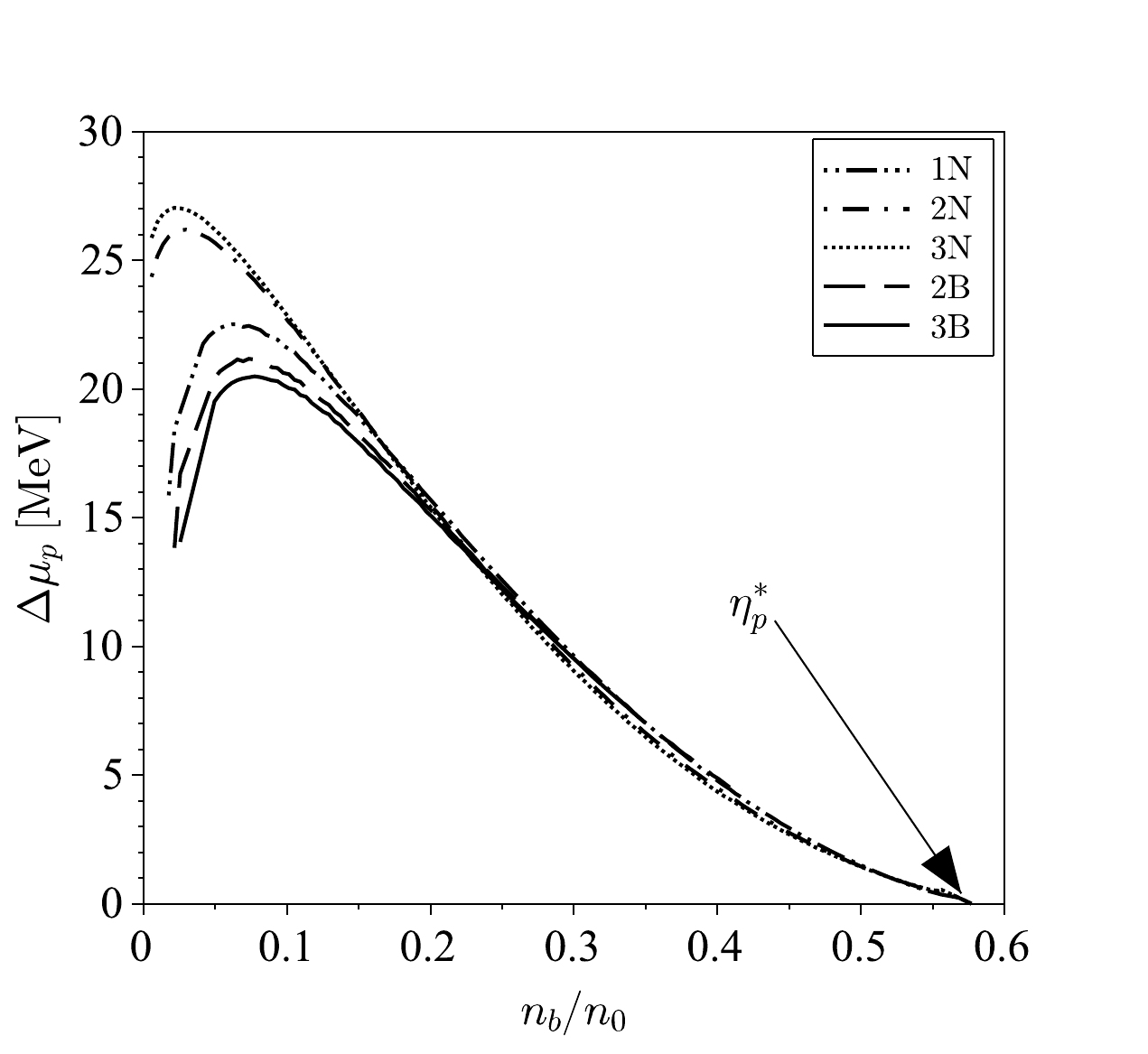}
\caption{The workfunction to transfer a proton from the nucleus to the surrounding pure neutron matter, Eq. (\ref{def_deltaMuP}), as function of
the mean baryon density $n_b$, calculated from the results shown in Figs. 3-6. The arrow marks the mean baryon density $\eta_p^{*}n_0$ corresponding
 to the proton drip.}
\end{figure}
The intersection of $\Delta\mu_p$ with the horizontal axis defines $\eta_p^{*}$, as shown schematically in Fig. 8 by the arrow.

\section{Discussion}
\subsection{General remarks}
It is instructive to compare the present results to the existing earlier evaluation of the surface tension, which is commonly provided by trial fitting functions $\sigma_s^{\rm fit}$ and $\sigma_c^{\rm fit}$.
In this case, the derivatives $(\sigma_s^{\rm fit})'$ and $(\sigma_c^{\rm fit})'$ are calculated analytically.
A standard choice is a pair of trial functions of the form
\begin{equation}\label{def_sigmaLimHolt}
  \sigma_s^{\rm fit}(x) = \sigma_0\frac{2^{\alpha+1} + q}{(1-x)^{-\alpha} + q + x^{-\alpha}}
\end{equation}
and
\begin{equation}\label{def_sigmaCfit}
  \sigma_c^{\rm fit}(x)= \frac{\sigma_{c0}}{\sigma_{0}}\sigma_s^{\rm fit}(x) \alpha_c(\beta_c - x).
\end{equation}

Let us start with the fitting $\sigma_s^{\rm fit}$ of the plain surface tension. The most relevant and up to date work is found in \cite{LimHolt2017}, where evaluation of $\sigma_s^{\rm fit}$ was based on an implicit fitting of the nucleus binding energy derived from the CLDM developed in \cite{LimHolt2017} to nuclear masses known from the experiment.
In this way, the parameters for the trial function in Eq. (\ref{def_sigmaLimHolt}) were inferred.
In Eq. (9) of \cite{LimHolt2017} the quantity $\sigma_s$ was represented by Eq. (\ref{def_sigmaLimHolt}) with
\begin{equation}\label{parametersLimHolt}
\sigma_0=1.186\; {\rm MeV\,fm}^{-2},\quad\alpha=3.4,\quad q=46.748.
\end{equation}

However, the nuclear masses in the framework of the CLDM are not the basic observables.
The basic observables in the context of neutron star matter are the stellar mass and radius, which are directly related to the parameters $n_0$, $B$, $K$, $S$, $L$.
Therefore, it is convenient to evaluate the nucleus surface properties in the CLDM by connecting them with $n_0$, $B$, $K$, $S$, $L$, which are in turn fitted within the CLDM to the terrestrial nuclear masses.
The robust connection is naturally achieved in the framework of the microscopic theoretical calculation of the surface properties such as the semi-infinite nuclear matter calculation in the Thomas-Fermi approximation leading to the data in Table III.

Figure 1 shows that $\sigma_s^{\rm fit}$ found in \cite{LimHolt2017} predicts a different set of numbers as compared to the values of $\sigma_s$ obtained here.
Specifically, it is clearly seen that the lines $\sigma_s$ and $\sigma_s^{\rm fit}$ in Fig. 1 tend to different numbers as $x\to1/2$.
Figures 1 and 2 make it obvious that the derivatives exhibit an even stronger disagreement with the data from Table III.
While the calculated values $\sigma_s'$ and $\sigma_c'$ are defined by the Skyrme model based on chosen values of $n_0$, $B$, $K$, $S$, $L$, the fitted values $(\sigma_s^{\rm fit})'$ and $(\sigma_c^{\rm fit})'$ are defined by the certain chosen form of the trial function in Eqs. (\ref{def_sigmaLimHolt}) and (\ref{def_sigmaCfit}).
It is true that the form of the fitting functions $\sigma_s^{\rm fit}$, $\sigma_c^{\rm fit}$ is not derived from microscopic physics, which is in contrast to the data for $\sigma_s$ and $\sigma_c$ displayed in Table III.

The novel implementation of evaluation of the surface properties performed in this paper has another significant advantage, which is the explicit consistency maintained between the derivation of the uniform part $\varepsilon$ and the surface part $\sigma_s$.
While in this paper the consistency is ensured by using the same underlying nucleon force for calculation of $\varepsilon$ and $\sigma_s$, this is not the case in the earlier work, where $\varepsilon$ was calculated from a nucleon force and $\sigma_s^{\rm fit}$ was evaluated from a completely different approach (the experimental mass fitting).
Actually, this argument is also relevant for evaluation of the curvature tension.

The curvature tension $\sigma_c$ was parametrized in \cite{NewtonGearheartLi2013} by Eq. (\ref{def_sigmaCfit}).
The parameter corresponding to $q$ in Eq. (\ref{def_sigmaLimHolt}) was denoted as $b$ and was not given explicitly in \cite{NewtonGearheartLi2013}.
Thus, I develop a fit of Eq. (\ref{def_sigmaCfit}) to the numerical data obtained in Table III, finding $\sigma_{c0}=0.4612$ MeV fm$^{-1}$, $\alpha_c=3.9027$ and $\beta_c=0.7562$.

A notable difference between $\sigma_c$ and $\sigma_c^{\rm fit}$ is also obvious in Fig. 2.
The curvature tension $\sigma_c$ was evaluated in \cite{NewtonGearheartLi2013}, where results of various authors were collected and their numerical variations were studied.
However, no constraint on a consistency between the underlying models for calculation of $\varepsilon$, $\sigma_s$ and $\sigma_c$ was imposed in \cite{NewtonGearheartLi2013}.
It is clear that respecting the constraint on the model self-consistency reduces the theoretical uncertainty.
This reduction is generally important for the best possible accuracy in theoretical study of neutron stars \cite{ChenPiekarewicz2014,Tews2017}.

Let us turn to the astrophysical implications of the present quantitative predictions based on the CLDM.
The results shown in Table I and Fig. 7 imply that the effect of $\sigma_c$ is to completely remove the bubble pasta phases from the ground state of the inner crust.
However, the bubble phases can be still expected in the inner crust thanks to the thermal fluctuations, as show the color dots in Fig. 7.
The ground state of the inner crust discussed above might be expected at absolute zero temperature, but at realistic temperature $T$ of observable neutron star matter it is meaningful to compare the energy gap per particle between the ground state and the excited states to the thermal energy per particle $k_BT$.
Inserting the solutions shown in Figs. 3-6 into the expressions for the energy density, Eqs. (\ref{wtotSect1}) and (\ref{def_wtot_BUB}), I
calculate the total internal energy density for each pasta phase.
Next, I determine the ground-state configuration at zero temperature, which is the one with the lowest internal energy density at fixed total baryon number and fixed temperature.
Notice that for fixed $n_b$, as in the present approach, the picture in terms of the energy per baryon is equivalent, up to a constant scaling factor, to the picture in terms of the energy density.
As a next step, I find all the possible structures whose energy per baryon differs from the ground state by less than the thermal energy $k_BT$.
The result of this comparison is shown in Fig. 7 by color dots, which represent the phases of pasta with energy per baryon not exceeding the ground state energy by more than $k_BT$, for three different temperatures, which are typical in the inner crust.

It is worth noting that to find the most thermodynamically favorable state of the system, one needs to minimize the thermodynamic potential $\Phi$.
For this purpose, it is necessary to define a set of thermodynamic variables, which imposes restrictions on the form of $\Phi$.
For instance, in this work the thermodynamic variables are the total number of baryons $N_b$ in the unit cell with volume $V_c$ and temperature $T$.
In this case, $\Phi(N_b,T)=E_{\rm tot}(N_b)-k_BTS(N_b,T)V_c$, where $E_{\rm tot}(N_b)$ is the internal energy per unit cell and $S(N_b,T)$ is the total entropy of the system per unit cell evaluated at given $N_b$ and $T$.
Alternatively, an equivalent description can be given in terms of the total pressure $P$ and temperature $T$ and in this case $\Phi(P,T)=E_{\rm tot}(P) -k_BTS(P,T)V_c + PV_c$, where the internal energy and entropy are calculated for given $P$ and $T$.
However, temperatures in the range of $10^8$--$10^9$ K are very small compared to the typical nuclear energy scale of 1 MeV, thus,
the relevant quantity is the internal energy at $T=0$ and the entropy can be neglected.

The energy differences are very sensitive to the details of the nucleon interaction and of the curvature tension.
It is known from the material science that the process of crystallization might lead to simultaneous formation of various crystal structures with close values of the thermodynamic potential, ending up with a polymorphic rather than a pure crystal structure.
A similar scenario for the crust was also anticipated in \cite{XiaEtal2021,NewtonEtal2022}.
For instance, \cite{NewtonEtal2022} was specifically aimed at study of the thermal fluctuation properties of the inner crust matter, where a convenient method to include various geometries of the unit cell into the Hartree-Fock scheme was suggested.
It is instructive to notice though, that the derived quantitative results of \cite{NewtonEtal2022} guarantee a theoretical uncertainty exceeding that of the previous results in the CLDM, which used the trial fitting functions for the surface part of the nucleus energy and was based on the approximate method of inclusion of the curvature \cite{NewtonGearheartLi2013}, because the Hartree-Fock scheme featured the lack of the spin-orbit coupling as shows Eq. (1) of \cite{NewtonEtal2022}, which is one of the essential ingredients in the nucleon force.
In the present paper, the spin-orbit coupling is taken into account when the nucleus energy is computed from Eq. (\ref{eqA11}).
To study the thermal fluctuation properties in \cite{NewtonEtal2022}, a free-energy-landscape was calculated from the simplified Hartree-Fock scheme based on the CLDM equilibrium states and the expected pasta phases were inferred from its local minima.
It would be instructive to evaluate sensitivity of the predicted energy-landscape studied in \cite{NewtonEtal2022} to the details of the nucleon force.

In this paper a more straightforward way to study the thermal fluctuations was used: Calculating the spectrum of the equilibrium configurations from the CLDM and comparing the spacings between the levels to the thermal energy per baryon.
This simple scheme leads to a set of levels (pasta phases) as function of mean baryon density, while respecting the self-consistency of the model and minimizing the theoretical uncertainty as much as possible within the CLDM.
\subsection{How does this work add to the existing knowledge?}
\subsubsection{Knowledge about the inner crust structure}
As a matter of fact, the previous knowledge about the inner crust structure has not been conclusive.
In this respect, earlier, much valuable work has been done on investigation of possible scenarios and outcomes.
Without an intention to present the complete list of the earlier works, I will mention what looks to me as the most characteristic papers.
The inner crust structure was studied using various approaches such as
the CLDM \cite{RavenhallPethickWilson1983,Hashimoto1984,LimHolt2017,ParmarEtAl2023},
the quantum molecular dynamics \cite{MaruyamaEtal1998,WatanabeEtAl2003,WatanabeEtAl2009},
the Thomas-Fermi method with the underlying equation of state obtained on the basis of either the Skyrme-type energy functionals \cite{WilliamsKoonin1985,Chamel2023} or the relativistic mean-field nuclear interactions \cite{AvanciniEtal2008} or the Brueckner-Hartree-Fock method \cite{SharmaEtal2015},
the Hartree-Fock approximation \cite{NewtonStone2009},
the density-functional theory \cite{SchuetrumpfEtal2019}.
Still, the results on appearance of the pasta phases have shown a strong model-dependence (for instance, see Fig. 1 in \cite{ParmarEtAl2023} and Fig. 4 in \cite{Chamel2023}).
Although the prediction of this paper (disappearance of the bubble-phase pasta from the ground state) is at odds with some of the earlier works (which have found the bubble-phase pasta in the ground state), but it is indeed compatible with the general trend of the model-dependent character of the present knowledge of the ground-state structure of the inner crust.

\subsubsection{Knowledge about the CLDM}
This work adds to the knowledge about the CLDM the new method to include the curvature tension using the new direct analytical result on the nucleus radius, Eqs. (\ref{solution_rN}) and (\ref{solution_rN_BUB}).
The new method makes it possible to judge analytically whether a given set of parameters is compatible with the CLDM equilibrium.
In particular, the CLDM equilibrium exists for a given set of parameters only if the nucleus radius calculated from Eqs. (\ref{solution_rN}) and (\ref{solution_rN_BUB}) is purely real-valued.

\subsubsection{When do the complex-valued $r_N$ or $r_B$ arise?}
To find out when do the complex-valued $r_N$ or $r_B$ arise, let us examine Eqs. (\ref{solution_rN}) and (\ref{solution_rN_BUB}).
The quantities $p$ in Eq. (\ref{p}) and $p^{\rm bub}$ in Eq. (\ref{p_BUB}) are always real-valued because ${q^{\rm bub}}^4>{r^{\rm bub}}^3$ and we always take the real-valued branch of the cubic root and forget about the remaining two complex-valued branches.
There is no ambiguity in taking one of the four roots in Eqs. (\ref{solution_rN}) and (\ref{solution_rN_BUB}), as explains Eq. (\ref{X}).
The only reason for appearance of the complex-valued roots is the second term on the right-hand side in Eq. (\ref{X}).
In the context of the bubble-phase pasta, $r_B$ might become complex-valued when $s<0$, where
\begin{equation}\label{def_s}
  s=q^{\rm bub}\left(\frac{2}{p^{\rm bub}}\right)^{3/2}-1.
\end{equation}

\subsubsection{What limits the range of $n_b/n_0$?}
There are two reasons why the range of $n_b/n_0$ in which the bubble-phase pasta is found is limited.
The first is that the intersection point of the curves $n_{no}^{P}$ and $n_{no}^{\mu}$ escapes.
To reveal this clearly, let us focus on the 2B phase.

The 2B phase was predicted in Cases 2 and 3 in vicinity of $n_b/n_0=0.54$ (see Table I).
In Case 4, the range of $n_b/n_0$ in which the 2B phase is found, can be seen in the Figures 3-5 and it is located at $n_b/n_0<0.4$.
Figure 9 shows the functions $n_{no}^{P}$ and $n_{no}^{\mu}$ at three different values of $n_b/n_0$.
The largest value $n_b/n_0=0.4085$ corresponds to the abscissa of the right-most point of the curves corresponding to 2B in Figs. 3-5.
The other two values of $n_b/n_0$ are shown to reveal the trend of the intersection of $n_{no}^P(n)$ and $n_{no}^\mu(n)$:
As the mean baryon density $n_b/n_0$ is increased, the intersection point moves toward the left and eventually escapes from the finite range of $n_{no}^{P}$ and $n_{no}^{\mu}$.
\begin{figure}
\includegraphics[width=3.5in]{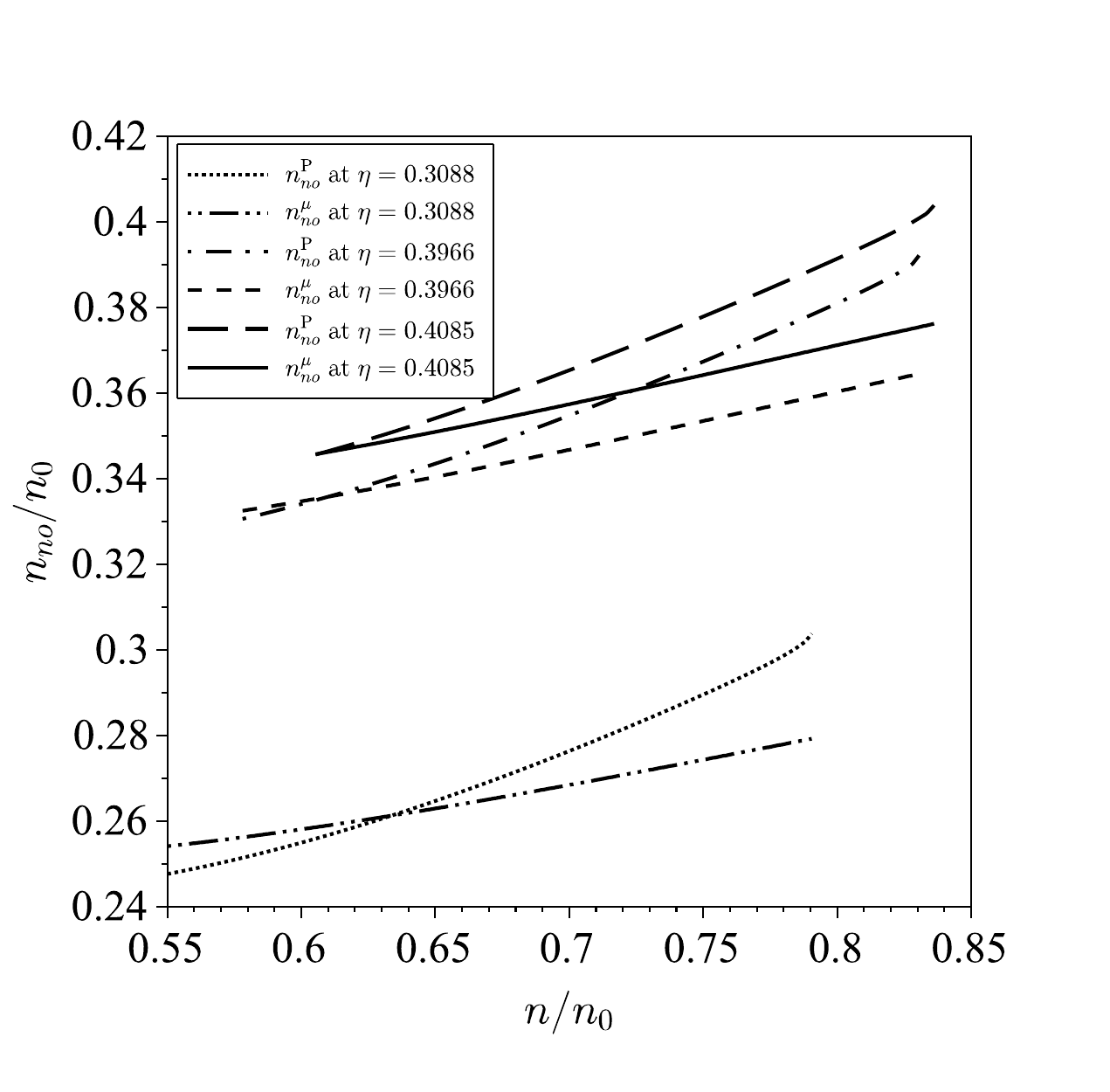}
\caption{The functions $n_{no}^{P}$ and $n_{no}^{\mu}$ at three different values of $n_b/n_0$. At each value of $n_b/n_0$, the intersection of $n_{no}^{P}$ and $n_{no}^{\mu}$ defines the CLDM equilibrium.}
\end{figure}

As the intersection point escapes from the domain in the $(n_{no}, n)$ plane, it is reasonable to ask, what is the factor that limits the domain in which $n_{no}^{P}$ and $n_{no}^{\mu}$ exist?
The short answer is that at values of $(n_{no}, n)$ beyond that range, the bubble radius becomes complex-valued and therefore the CLDM equilibrium does not exist.
The detailed mechanism of the domain limitation can be revealed by considering some certain values.

For instance, consider $(n_{no}/n_0=0.1634,\; n/n_0=0.5500)$ which lie beyond the domain (at $n_b/n_0=0.4085$).
Note that the domain can be easily seen from the corresponding curves in Fig. 9.
The function $n_{no}=n_{no}^{P}(n)$ ensures that $P_i^{\rm bub} = P_o^{\rm bub}$.
In order to compute $P_i^{\rm bub}$ and $P_o^{\rm bub}$, one must use the values of $x$ which satisfy the beta-equilibrium condition, Eq. (\ref{var_2}), for the chosen values of $n_b$, $n_{no}$ and $n$.
The equilibrium value of $x$, which for convenience I denote as $x=x_\beta$ is sought in the range $0<x<0.5$.
For each of the $x$ value, the quantity $r_B(x)$ is calculated and the range of $x$, which generates complex-valued $r_B(x)$ is discarded.
The remaining values of $x$ generate real-valued $r_B(x)$, which are inserted into Eq. (\ref{var_2}).
If none of the real-valued $r_B(x)$ satisfy Eq. (\ref{var_2}) (in other words, if $x_\beta$ does not lie in the interval of $x$ for which $s\geq0$ is satisfied), that implies the CLDM equilibrium does not exist for the chosen values of $n_b$, $n_{no}$ and $n$.

A concrete example of the situation described above is shown in Fig. 10, which displays the quantity $s(x)$.
The values of $(n_{no}, n)$ are chosen beyond the existence domain of the CLDM equilibrium for the given $n_b$.
Consequently, the beta-equilibrium, Eq. (\ref{var_2}), can not be satisfied in the physically meaningful interval $s\geq0$.
To verify the model-independence of this numerical mechanism, the quantity $s(x)$ was also calculated using the empirical fits to the surface and curvature tensions, Eqs. (\ref{def_sigmaLimHolt}) and (\ref{def_sigmaCfit}).
The result is shown in Fig. 10 by the dotted lines.
\begin{figure}
\includegraphics[width=3.5in]{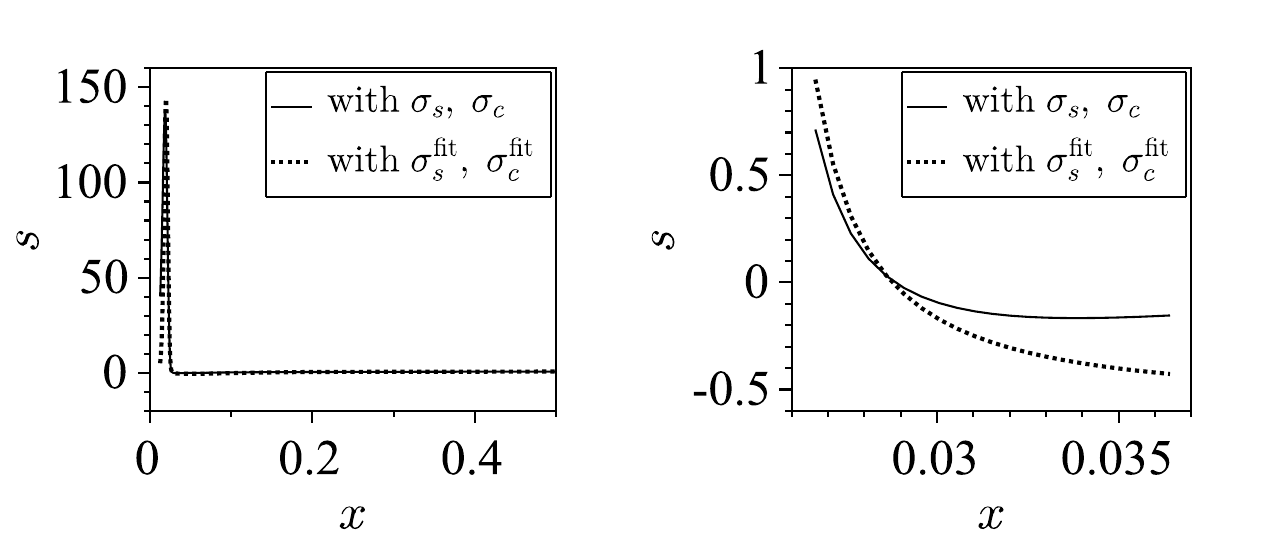}
\caption{The quantity $s$, Eq. (\ref{def_s}), computed with $(n_{no}/n_0=0.1634,\; n/n_0=0.5500)$ and $n_b/n_0=0.4085$. The surface and curvature tensions were calculated from Table III (solid lines) and from Eqs. (\ref{def_sigmaLimHolt}) and (\ref{def_sigmaCfit}) (dashed lines). Left panel: The entire domain of the function. Right panel: enlarged view in vicinity of the point at which $s(x)$ becomes negative.}
\end{figure}

\section{Conclusion}
In this paper we have carefully incorporated the curvature tension into the CLDM using a novel analytical step in the solution scheme.
In particular, the curvature has been included by the exact algebraic solution to the CLDM equation that determines the nucleus radius.
The full set of the CLDM equations has been solved using a novel graphical method.
These improvements have provided a significant impact on the final solution, as non-physical solutions are sorted out analytically and immediately discarded.
The new graphical method completely circumvents the convergence and the initial guess problems routinely encountered in the usual numerical schemes such as the Newton method and the steepest-descent method.

Another improvement, which have proven to cause minor but still important impact is an explicit maintaining of self-consistency of the numerical input to the CLDM.
The surface part of the nuclear interaction has been included on the same basis as the bulk part, which is matched to the equation of state of infinite symmetric nuclear matter defined mainly by the parameters $n_0$, $B$, $K$, $S$, $L$.
In contrast, in the earlier work the surface part of the nuclear interaction was represented by trial functions, which were found by matching predictions of the CLDM to measured masses of the atomic nucleus independently of matching of the bulk part, thereby compromising self-consistency of the CLDM and increasing the theoretical uncertainty.

The numerical results of this paper are based on the equation of state represented by the Skyrme parametrization (Sk$\chi$450) and constrained by the effective chiral field theory following \cite{LimHolt2017}. 
The ground-state structure of the inner crust is predicted to be composed exclusively of 3N, 2N and 1N pasta phases.
This is likely a model-dependent conclusion, but probing the predictions with other parametrizations goes beyond the scope of this paper.

The results reported as Cases 2 and 3 (Table I) demonstrate, firstly, that the present implementation, in particular the novel graphical method, is capable of reproducing the earlier results and, secondly, that the replacement of the fit to $\sigma_s$ by the self-consistent data while keeping $\sigma_c=0$ does not lead to a dramatic modification of the range of $n_b$ in which the bubble-phase pasta are found.
The latter occurs only due to inclusion of $\sigma_c\neq0$.

The numerical results on the CLDM equilibrium states suggest that at a typical temperature of $10^8$ K expected in the inner crust, the matter is likely a thermodynamic mixture of various pasta phases, in almost the entire relevant range of the mean baryon density $n_b$, except for the low-density edge, where 3N pasta phase is predicted.
Though  in principle, the pure pasta phases in the inner crust are not ruled out.

A model-independent conclusion that follows from the numerical results reported in this work is that even with more precise results on the ground state structure of the inner crust, the realistic structure of neutron star matter remains obscured by thermodynamic fluctuations, because the difference of energies of the various pasta phases is smaller that the thermal energy per particle at $10^8$ K.
Various options for the inner crust structure will be within the systematic theoretical uncertainty of not only the CLDM but also of more sophisticated approaches such as the Hartree-Fock-Bogoliubov calculations, as long as the systematic uncertainty of the thermal fluctuation properties of the inner crust structure is not under control.
Although this conclusion is not new and has been known from the earlier studies, this point is worth to emphasize again.

This work also offers a novel theoretical tool in the context of the CLDM.
The new method of computation of the nucleus $r_N$ and the bubble $r_B$ radius, Eqs. (\ref{solution_rN}) and (\ref{solution_rN_BUB}), makes it possible to judge analytically whether a given set of parameters $n_b$, $n_{no}$, $n$ and $x$ satisfy the CLDM equilibrium as defined in Eqs. (\ref{var_1})-(\ref{var_4}).
The analytical condition for the existence of the CLDM equilibrium is that $r_N$ and $r_B$ are real-valued.
The reason of narrowing of the range of $n_b$ in which the CLDM equilibrium exists as a result of inclusion of $\sigma_c\neq0$, as seen from the curves corresponding to 2B and 3B in Fig. 3-5, is that beyond this range $r_N$ and $r_B$ are complex-valued and do not represent physically meaningful solutions.

\section*{Acknowledgements}
I thank Xavier Vi\~nas for providing the numerical data on the nuclear surface properties and for illuminating discussions.
I am also grateful to Gentaro Watanabe for useful remarks.

\appendix
\section{Properties of uniform nuclear matter}
Table II displays properties of uniform nuclear matter defined by the parameters of the nucleon force $t_0$, $t_1$, $t_2$, $t_3$, $t_4$, $x_0$, $x_1$, $x_2$, $x_3$, $x_4$, $\alpha_1$, $\alpha_2$, which are constrained by the empirical nuclear parameters $n_0$, $B$, $K$, $S$, $L$ defined in Eqs.
(\ref{def_B})-(\ref{def_K}).
The nuclear symmetric saturation density $n_0$ is fine-tuned so that the empirical condition for the pressure inside the nucleus, Eq. (\ref{PnucSaturation}), is satisfied to a good precision.
\begin{table}
\begin{tabular}{|c|c|}
  \hline
  $t_0\;[{\rm MeV\,fm}^{3}]$ & -1803.2928 \\
  \hline
  $t_1\;[{\rm MeV\,fm}^{5}]$ &  301.8208 \\
  \hline
  $t_2\;[{\rm MeV\,fm}^{5}]$ &  -273.2827 \\
  \hline
  $t_3\;[{\rm MeV\,fm}^{3(\alpha_1+1)}]$ &  12783.8619 \\
  \hline
  $t_4\;[{\rm MeV\,fm}^{3(\alpha_2+1)}]$ &  564.1049 \\
  \hline
  $x_0$ &  0.4430 \\
  \hline
  $x_1$ &  -0.3622 \\
  \hline
  $x_2$ &  -0.4105 \\
  \hline
  $x_3$ &  0.6545 \\
  \hline
  $x_4$ &  -11.3160 \\
  \hline
  $\alpha_1$ &  1/3 \\
  \hline
  $\alpha_2$ &  1 \\
  \hline
  $n_0\;[{\rm fm}^{-3}]$ & 0.1561 \\
  \hline
  $P_{\rm nuc}\;[{\rm MeV\,fm}^{-3}]$ & $3\times10^{-6}$ \\
  \hline
  $B\;[{\rm MeV}]$ & -15.91 \\
  \hline
  $K\;[{\rm MeV}]$ & 239.3 \\
  \hline
  $S\;[{\rm MeV}]$ & 31.44 \\
  \hline
  $L\;[{\rm MeV}]$ & 42.07 \\
  \hline
\end{tabular}
\caption{Nuclear interaction parameters $t_0$, $t_1$, $t_2$, $t_3$, $t_4$, $x_0$, $x_1$, $x_2$, $x_3$, $x_4$, $\alpha_1$ and $\alpha_2$ adopted from \cite{LimHolt2017}, leading to the empirical nuclear parameters $B$, $K$, $S$, $L$ calculated from Eqs. (\ref{def_B})-(\ref{def_K}). The symmetric nuclear density $n_0$ is fine-tuned to satisfy the empirical condition given in Eq. (\ref{PnucSaturation}).}
\end{table}
\section{Properties of nuclear surface}
Table III shows our results obtained from Eqs. (\ref{SurTen1}) and (\ref{eqA7}) for the nuclear surface tension $\sigma_s$ in units of ${\rm MeV\,fm}^{-2}$ and for the curvature tension $\sigma_c$ in units of ${\rm MeV\,fm}^{-1}$ and their derivatives with respect to $x$.
\begin{table}
\begin{tabular}{|c|c|c|c|c|}
  \hline
  $x$ & $\sigma_s\,[{\rm MeV fm}^{-2}]$ & $\partial\sigma_s/\partial x$ & $\sigma_c\,[{\rm MeV fm}^{-1}]$ & $\partial\sigma_c/\partial x$ \\
  \hline
  0.0250 & 1.23$\times10^{-4}$ & 0.01154 & -1.0$\times10^{-4}$ & 0.1258  \\
  \hline
0.0375 & 9.91$\times10^{-4}$ & 0.1304 & 0.001521 & 0.1875 \\
  \hline
0.0500 & 0.003427 & 0.2598 & 0.005366 & 0.4360 \\
  \hline
0.0625 & 0.007467 & 0.3802 & 0.01206 & 0.5526 \\
  \hline
0.0750 & 0.01295 & 0.5053 & 0.01880 & 0.5502 \\
  \hline
0.0875 & 0.02018 & 0.6520 & 0.02606 & 0.6368 \\
  \hline
0.1000 & 0.02934 & 0.8179 & 0.03489 & 0.7670 \\
  \hline
0.1125 & 0.04073 & 1.0079 & 0.04528 & 0.8994 \\
  \hline
0.1250 & 0.05463 & 1.2210 & 0.05746 & 1.0527 \\
  \hline
0.1375 & 0.07135 & 1.4583 & 0.07168 & 1.2267 \\
  \hline
0.1500 & 0.09119 & 1.7198 & 0.08822 & 1.4249 \\
  \hline
0.1625 & 0.1144 & 2.0058 & 0.1074 & 1.6380 \\
  \hline
0.1750 & 0.1414 & 2.3133 & 0.1292 & 1.8672 \\
  \hline
0.1875 & 0.1723 & 2.6371 & 0.1541 & 2.1066 \\
  \hline
0.2000 & 0.2074 & 2.9697 & 0.1819 & 2.3434 \\
  \hline
0.2125 & 0.2466 & 3.3007 & 0.2126 & 2.5542 \\
  \hline
0.2250 & 0.2898 & 3.6179 & 0.2456 & 2.7174 \\
  \hline
0.2375 & 0.3369 & 3.9049 & 0.2803 & 2.8424\\
  \hline
0.2500 & 0.3873 & 4.1527 & 0.3163 & 2.8970 \\
  \hline
0.2625 & 0.4405 & 4.3513 & 0.3525 & 2.8886  \\
  \hline
0.2750 & 0.4959 & 4.4946 & 0.3883 & 2.8267  \\
  \hline
0.2875 & 0.5526 & 4.5808 & 0.4229 & 2.7054  \\
  \hline
0.3000 & 0.6101 & 4.6061 & 0.4558 & 2.5453 \\
  \hline
0.3125 & 0.6675 & 4.5678 & 0.4864 & 2.3485 \\
  \hline
0.3250 & 0.7240 & 4.4627 & 0.5143 & 2.1193 \\
  \hline
0.3375 & 0.7786 & 4.2349 & 0.5386 & 1.6866 \\
  \hline
0.3500 & 0.8294 & 3.8600 & 0.5555 & 0.9633 \\
  \hline
0.3625 & 0.8750 & 3.4700 & 0.5631 & 0.3690 \\
  \hline
0.3750 & 0.9163 & 3.1317 & 0.5657 & 0.04603 \\
  \hline
0.3875 & 0.9534 & 2.8083 & 0.5646 & -0.2078 \\
  \hline
0.4000 & 0.9865 & 2.4923 & 0.5607 & -0.4167 \\
  \hline
0.4125 & 1.0157 & 2.1788 & 0.5543 & -0.5977 \\
  \hline
0.4250 & 1.0410 & 1.8671 & 0.5458 & -0.7558 \\
  \hline
0.4375 & 1.0624 & 1.5563 & 0.5355 & -0.8968\\
  \hline
0.4500 & 1.0799 & 1.2455 & 0.5235 & -1.0245 \\
  \hline
0.4625 & 1.0935 & 0.9344 & 0.5099 & -1.1417\\
  \hline
0.4750 & 1.1033 & 0.6232 & 0.4949 & -1.2502\\
  \hline
0.4875 & 1.1091 & 0.3116 & 0.4787 & -1.3516 \\
  \hline
0.5000 & 1.1110 & -1.2$\times10^{-4}$ & 0.4612 & -1.4472\\
  \hline
\end{tabular}
\caption{The plain surface tension $\sigma_s(x)$ and the curvature tension $\sigma_c(x)$ computed from Eqs. (\ref{SurTen1}) and (\ref{eqA7}).}
\end{table}

\section{Nucleus phases in liquid drop model}
It is convenient to derive the equations for the nucleus phases and for the bubble phases in separate sections.
The nucleus volume is equal to
\begin{equation}\label{def_VN}
  V_N=\left\{ \begin{array}{c}
                ({4}/{3}) \pi r_N^3,\quad (3N) \\
                \pi r_N^2\times L,\quad (2N) \\
                2r_N\times L^2,\quad (1N)
              \end{array}
   \right.
\end{equation}
where $L\rightarrow\infty$ is the length of the rod-like nucleus and $L^2\rightarrow\infty$ is the area of the slab-like nucleus.
The limit $L\rightarrow\infty$ should be understood as large enough $L$ so that the Coulomb energy per unit length in 2N and 2B phases or area in 1N phase does not depend on $L$.

Let us consider each of the variational equations, Eq. (\ref{def_VarEq}), for the set of variables introduced in Eq. (\ref{VarSet2}).

\subsection{The nuclear energy law}
We start with the optimization of the nucleus radius with the other variables in Eq. (\ref{VarSet2}) kept fixed, corresponding to $Y=r_N$ in Eq. (\ref{def_VarEq}).
Notice that since $n_{ni}=(1-x)n$, we actually have to fix $n$ rather than $n_{ni}$.
This yields
\begin{equation}\label{var_eq_rN}
  \left.\frac{\partial(w_{\rm tot})}{\partial r_N}\right|_{x,n,n_{n0},u}=0.
\end{equation}
The resulting equation combines the surface energy of the plain interface, the curvature energy and the Coulomb energy of the nucleus:
\begin{equation}\label{ws_2wCL}
  w_{\rm s} + 2w_{\rm cur}=2w_{\rm C+L}.
\end{equation}

\subsection{Chemical equilibrium}
Next, we find the optimal proton fraction inside the nucleus corresponding to $Y=x$ in Eq. (\ref{def_VarEq}) and the other variables in Eq. (\ref{VarSet2}) kept fixed.
Again, we account for $n_{ni}=(1-x)n$ and fix $n$ in the calculation, yielding
\begin{equation}\label{var_eq_x}
  \left.\frac{\partial(w_{\rm tot})}{\partial x}\right|_{r_N,n,n_{no},u}=0.
\end{equation}
Working out the derivative we find
\begin{eqnarray}
 &&  \label{betaEquilCond}
  \mu_{e}+(m_p-m_n)c^2 = -\left.\frac{\partial \varepsilon}{\partial x}\right|_{n} \\
\nonumber  &&  - \frac{1}{un}\left.\frac{\partial (w_{\mathrm{s}} + w_{\mathrm{C+L}} + w_{\mathrm{cur}})}{\partial x}\right|_{r_N,n,n_{no},u}.
\end{eqnarray}
The electron chemical potential $\mu_{e}$ is:
\begin{equation}\label{def_mu_cluster}
  \mu_{e}=\frac{1}{un}\left.\frac{\partial w_{\mathrm{e}}}{\partial x}\right|_{r_N,n,n_{no},u}=\hbar c(3\pi^2uxn)^{\frac{1}{3}}.
\end{equation}
Equation (\ref{betaEquilCond}) may be represented as the chemical equilibrium condition:
\begin{equation}\label{betaEquilCond_form}
  \mu_{e}=\mu_{ni} - \mu_{pi},
\end{equation}
where the neutron and the proton chemical potentials are given below by Eqs. (\ref{Muni}) and (\ref{Mupi}), respectively.

\subsection{Continuity of neutron chemical potential across the nucleus interface}
In the next step we find the optimal number density of neutrons inside the nucleus, corresponding to $Y=n_{ni}$ in Eq. (\ref{def_VarEq}) and the other variables in Eq. (\ref{VarSet2}) kept fixed.
It is convenient to fix the nucleus radius $r_N$, the total number of neutrons in the cell $N_n$, the proton number density in the nucleus $n_{pi}$ and $u$:
\begin{equation}\label{var_nni}
  \left.\frac{\partial(w_{\rm tot})}{\partial n_{ni}}\right|_{N_{n},n_{pi},r_N,u}=0.
\end{equation}
When $n_{ni}$ is optimized, the energy to add a neutron to the nucleus is equal to the energy to add a neutron to the dripped neutrons, $\mu_{ni}=\mu_{no}$, which can be explicitly seen as following.

First, notice that fixing of $N_{n}$, $n_{pi}$, $r_N$, $u$ induces the following relations:
\begin{eqnarray}
\label{deltax_deltan}&& n\delta x + x\delta n=0,\\
\label{deltanni_deltanno}&& u\delta n_{ni} + (1-u)\delta n_{no}=0.
\end{eqnarray}
Then the partial derivative reads
\begin{eqnarray}\label{partial_nni_initial}
 &&  \left.\frac{\partial}{\partial n_{ni}}\right|_{N_{n},n_{pi},r_N,u}= \\
  \nonumber && \left.\frac{\partial}{\partial n}\right|_{x,n_{no},r_N,u} + \frac{\delta n}{\delta x}\left.\frac{\partial}{\partial x}\right|_{n,n_{no},r_N,u} + \frac{\delta n_{no}}{\delta n_{ni}}\left.\frac{\partial}{\partial n_{no}}\right|_{n,x,r_N,u}.
\end{eqnarray}
Using Eqs. (\ref{deltax_deltan}) and (\ref{deltanni_deltanno}), we find
\begin{eqnarray}\label{partial_nni}
 &&  \left.\frac{\partial}{\partial n_{ni}}\right|_{N_{n},n_{pi},r_N,u}= \\
  \nonumber && \left.\frac{\partial}{\partial n}\right|_{x,\,n_{no},r_N,u} - \frac{x}{n}\left.\frac{\partial}{\partial x}\right|_{n,n_{no},r_N,u} - \frac{u}{1-u}\left.\frac{\partial}{\partial n_{no}}\right|_{n,x,r_N,u}.
\end{eqnarray}
Application of these partial derivatives in Eq. (\ref{var_nni}) yields
\begin{equation}\label{continuityMu}
  \mu_{ni}=\mu_{no},
\end{equation}
where
\begin{eqnarray}
\nonumber   \mu_{ni}=m_nc^2 + \left(\left.\frac{\partial}{\partial n}\right|_{x} - \frac{x}{n}\left.\frac{\partial}{\partial x}\right|_{n}\right)\left[n\varepsilon(n,x)\right] \\
\label{Muni}   - \frac{x}{un} \left.\frac{\partial  (w_{\rm s}+w_{\rm cur})}{\partial x}\right|_{r_N,u},
\end{eqnarray}
and
\begin{equation}
\label{Muno}  \mu_{no}=\frac{1}{1-u}\left.\frac{\partial  w_{\mathrm{no}}}{\partial n_{no}}\right|_{u} = m_nc^2 + \frac{\partial }{\partial n_{no}}[n_{no}\varepsilon(n_{no},0)].
\end{equation}
Notice that Eqs. (\ref{Muni}) and (\ref{Muno}) do not contain the contribution from $w_{\rm C+L}$ by virtue of Eq. (\ref{deltax_deltan}).

The derivatives within $\mu_{ni}$ are calculated as following:
\begin{eqnarray}
\label{dws_dx} && -\frac{x}{un}\frac{\partial  w_{\rm s}}{\partial x}=-\frac{xd}{nr_N}\frac{\partial  \sigma_s(x)}{\partial x},\\
\label{dwC_dx} && -\frac{x}{un}\frac{\partial  w_{\rm cur}}{\partial x}=-\frac{xd(d-1)}{nr_N^2}\frac{\partial  \sigma_c(x)}{\partial x}.
\end{eqnarray}

By combining Eqs. (\ref{betaEquilCond}), (\ref{betaEquilCond_form}) and (\ref{Muni}) we find the proton chemical potential inside the nucleus including the surface and Coulomb corrections:
\begin{eqnarray}
\nonumber   &&\mu_{pi}=m_pc^2 + \left(\left.\frac{\partial}{\partial n}\right|_{x} + \frac{1-x}{n}\left.\frac{\partial}{\partial x}\right|_{n}\right)\left[n\varepsilon(n,x)\right] \\
\label{Mupi}  &&+ \frac{1-x}{un} \left.\frac{\partial  (w_{\rm s}+w_{\rm cur})}{\partial x}\right|_{r_N,u} + \frac{1}{un} \left.\frac{\partial  (w_{\rm C+L})}{\partial x}\right|_{n,r_N,u}.
\end{eqnarray}

The proton chemical potential outside the nucleus can be found directly from the bulk properties of the pure neutron matter:
\begin{equation}
\label{Mupo}  \mu_{po}=\left.\left[\left(\left.\frac{\partial}{\partial n}\right|_{x} + \frac{1}{n}\left.\frac{\partial}{\partial x}\right|_{n}\right)\left[n\varepsilon(n,x)\right]\right]\right|_{n=n_{no},\,x=0}.
\end{equation}

\subsection{Continuity of pressure across the nucleus interface}
Finally, we find the optimal nucleus fraction, corresponding to $Y=u$ in Eq. (\ref{def_VarEq}) and the other variables in Eq. (\ref{VarSet2}) kept fixed.
It is convenient to fix the total number of neutrons in the cell $N_n$, the number of neutrons outside the nucleus $N_{no}$, the protons fraction $x$ and the volume of the unit cell
\begin{equation}\label{var_u}
  \left.\frac{\partial(w_{\rm tot})}{\partial u}\right|_{N_{n},N_{no},x,V_c}=0.
\end{equation}
Optimization of $u$ implies that the pressure in the nucleus is equal to the pressure in the dripped neutrons, $P_{i}=P_{o}$.
The explicit form of this equality is obtained as following.

Fixing $N_{n}$, $N_{no}$, $x$ and $V_c$ induces the following relations:
\begin{eqnarray}
\label{deltau_deltan} &&  n\delta u + u\delta n=0,\\
\label{deltanno_deltau}&&  (1-u)\delta n_{no} - n_{no}\delta u=0.
\end{eqnarray}
Working out the partial derivative we find
\begin{eqnarray}\label{partial_u_initial}
 &&  \left.\frac{\partial}{\partial u}\right|_{N_{n},N_{no},x,V_c}= \\
  \nonumber && \left.\frac{\partial}{\partial u}\right|_{n,x,n_{no},V_c} + \frac{\delta n}{\delta u}\left.\frac{\partial}{\partial n}\right|_{n_{no},x,r_N,u} + \frac{\delta n_{no}}{\delta u}\left.\frac{\partial}{\partial n_{no}}\right|_{n,x,r_N,u}.
\end{eqnarray}
Using Eqs. (\ref{deltau_deltan}) and (\ref{deltanno_deltau}), we find
\begin{eqnarray}\label{partial_u}
 &&  \left.\frac{\partial}{\partial u}\right|_{N_{n},N_{no},x,V_c}= \\
  \nonumber && \left.\frac{\partial}{\partial u}\right|_{n,x,n_{no},V_c} - \frac{n}{u}\left.\frac{\partial}{\partial n}\right|_{n_{no},x,r_N,u} + \frac{n_{no}}{1-u}\left.\frac{\partial}{\partial n_{no}}\right|_{n,x,r_N,u}.
\end{eqnarray}
Application of these partial derivatives in Eq. (\ref{var_u}) yields the equality of the pressure in the nucleus and the pressure in the dripped neutrons
\begin{equation}\label{continuityP}
  P_{i}=P_{o},
\end{equation}
where
\begin{eqnarray}
\nonumber &&  P_{i} = n^2\left.\frac{\partial \varepsilon}{\partial n}\right|_{x} - \left(\left.\frac{\partial}{\partial u}\right|_{n,x,n_{no},V_c} - \frac{n}{u}\left.\frac{\partial}{\partial n}\right|_{n_{no},r_N,u}\right)\\
\label{Pi} && \times(w_{\rm s} + w_{\rm C+L}+w_{\rm cur}),
\end{eqnarray}
and
\begin{equation}
\label{Po} P_{o} = \frac{n_{no}}{1-u}\left.\frac{\partial w_{\mathrm{no}}}{\partial n_{no}}\right|_{u} - \frac{w_{\mathrm{no}}}{1-u} \\
= n_{no}^2\frac{\partial \varepsilon(n_{no},0)}{\partial n_{no}}.
\end{equation}
With the help of Eq. (\ref{deltau_deltan}) and the relations
\begin{eqnarray}
&& \left.\partial(u/r_N)/\partial u\right|_{V_c}=(d-1)/r_Nd,\\
&& \left.\partial(u/r_N^2)/\partial u\right|_{V_c}=(d-2)/r_N^2d,\\
&& \left.\partial(r_N^2)/\partial u\right|_{V_c}=2r_N^2/ud,
\end{eqnarray}
the derivatives within $P_{i}$ are calculated as following:
\begin{eqnarray}
\label{dws_du}&&\left.\frac{\partial w_{\rm s}}{\partial u}\right|_{N_{n},N_{no},x,V_c}=\frac{d-1}{r_N}\sigma_s(x),\\
\label{dwc_du}&&\left.\frac{\partial w_{\rm cur}}{\partial u}\right|_{N_{n},N_{no},x,V_c}=\frac{(d-2)(d-1)}{r_N^2}\sigma_c(x),
\end{eqnarray}
and
\begin{eqnarray}
  \nonumber  \left(\left.\frac{\partial}{\partial u}\right|_{n,x,n_{no},V_c} - \frac{n}{u}\left.\frac{\partial}{\partial n}\right|_{n_{no},r_N,u}\right)w_{\rm C+L}\\
  \label{dwCL_du} =2\pi (e x n r_N)^2 (u-1)g_d,
\end{eqnarray}
where $g_3=2/15$, $g_2=1/4$ and $g_1=2/3$.

\section{Bubble phases in liquid drop model}
Derivation of the variational equations in the bubble phases is analogous to the nucleus phases.
However, the basic definitions have to be carefully modified.
\subsection{Basic definitions}
Let us introduce the radius $r_B$ of the bubble.
Then the bubble volume is
\begin{equation}\label{def_VN_BUB}
  V_B=\left\{ \begin{array}{c}
                ({4}/{3}) \pi r_B^3,\quad (3B) \\
                \pi r_B^2\times L,\quad (2B),
              \end{array}
   \right.
\end{equation}
where we have taken into account the equivalence of the case of 1-dimensional nucleus and the 1-dimensional bubble and thus we will not consider the 1B phase.
Similarly to the nucleus phase we denote as $L\rightarrow\infty$ the length of the rod-like bubble and as $L^2\rightarrow\infty$ the area of the slab-like bubble.

The protons are uniformly distributed outside the bubble with the number density
\begin{equation}\label{def_np_BUB}
  n_p=\frac{N_p}{V_c-V_B},
\end{equation}
where $N_p$ is the number of protons outside the bubble.
The bubble volume fraction is
\begin{equation}\label{def_u_BUB}
  u^{\rm bub}=\frac{V_B}{V_c}=\left(\frac{r_B}{r_c}\right)^d=\frac{n-n_b}{n-n_{no}},
\end{equation}
where $r_c$ is the radius of the unit cell, $d=3$ for 3B and $d=2$ for 2B.
The number density of neutrons outside the bubble is
\begin{equation}\label{def_nni_BUB}
  n_{ni}=\frac{N_{ni}}{V_c-V_B}=(1-x)n,
\end{equation}
where $N_{ni}$ is the number of neutrons outside the bubble. The proton fraction, the baryon density outside the bubble and the number density of neutrons inside the bubble are defined as following:
\begin{equation}\label{def_x_BUB}
  x=\frac{N_{p}}{N_{p}+N_{ni}},\;n=\frac{N_{p}+N_{ni}}{V_c-V_B},\;n_{no}=\frac{N_{no}}{V_B}.
\end{equation}
From Eq. (\ref{def_u_BUB}), the mean baryon number density can be written as
\begin{equation}\label{def_nb_BUB}
  n_{b}=(1-u^{\rm bub})n+u^{\rm bub}n_{no}.
\end{equation}
The total energy density in the bubble phase is
\begin{equation}\label{def_wtot_BUB}
  w_{\mathrm{tot}}^{\rm bub}=\frac{E_{\mathrm {tot}}^{\rm bub}}{V_c}
\end{equation}
and it includes the contributions analogous to the nucleus phase, Eq. (\ref{wtotSect1}):
\begin{equation}\label{wtot_BUB}
w_{\mathrm{tot}}^{\rm bub}=w_{\mathrm{nuc}}^{\rm bub} + w_{\mathrm{s}}^{\rm bub} + w_{\mathrm{cur}}^{\rm bub} + w_{\mathrm{C+L}}^{\rm bub} + w_{\mathrm{no}}^{\rm bub} + w_{\mathrm{e}}^{\rm bub}.
\end{equation}
The contribution from the rest masses and from the uniform nuclear matter outside the bubble reads
\begin{equation}\label{def_wnuc_BUB}
  w_{\rm nuc}^{\rm bub}=(1-u^{\rm bub})n\left[\left(1-x\right)m_n + xm_p\right]c^2 + (1-u^{\rm bub})n\varepsilon(n,x),
\end{equation}
where $\varepsilon$ is the energy per baryon specified in Eq. (\ref{def_exn}).

The surface tension of the plain interface is contained in
\begin{equation}\label{def_ws_BUB}
  w_{\rm s}^{\rm bub}=\frac{u^{\rm bub}d}{r_B}\sigma_s(x),
\end{equation}
where $\sigma_s(x)$ is given in Table III.
The mean principal curvature of the bubble surface is $-2/r_B$ for 3B and $-1/r_B$ for bubbles of 2B.
Thus, the energy density associated with curvature of the bubble surface has the form
\begin{equation}\label{def_wbend_BUB}
  w_{\rm cur}^{\rm bub}=-\frac{u^{\rm bub}d(d-1)}{r_B^2}\sigma_c,
\end{equation}
where $\sigma_c(x)$ is given in Table III.
Notice that the surface curvature contribution for bubbles has the opposite sign as compared to that for nuclei.

The Coulomb energy density is
\begin{equation}\label{def_wCoul_BUB}
  w_{\rm C+L}^{\rm bub}=2\pi(enxr_B)^2u^{\rm bub}f_d(u^{\rm bub}),
\end{equation}
where the function $f_d$ is given in Eq. (\ref{def_fdu}).

The energy density of the dripped neutrons $w_{\mathrm{no}}^{\rm bub}$ is
\begin{equation}\label{def_wno_BUB}
  w_{\mathrm{no}}^{\rm bub}=u^{\rm bub}n_{no}\left[m_nc^2 + \varepsilon(n_{no},0)\right].
\end{equation}

The electron energy density $w_{\rm e}^{\rm bub}$ is
\begin{equation}\label{def_we_BUB}
  w_{\rm e}^{\rm bub}=\frac{3}{4}\hbar c(3\pi^2)^{1/3} \left[(1-u^{\rm bub})nx\right]^{4/3},
\end{equation}
where the electrical neutrality condition in the unit cell is ensured.
We proceed with the variational equations analogously to the case of the nucleus phases.

\subsection{The nuclear energy law}
From the equation
\begin{equation}\label{var_eq_rN_BUB}
  \left.\frac{\partial(w_{\rm tot}^{\rm bub})}{\partial r_B}\right|_{n,x,n_{no},u^{\rm bub}}=0
\end{equation}
we obtain the nuclear energy law
\begin{equation}\label{ws_2wCL_BUB}
  w_{\rm s}^{\rm bub} + 2w_{\rm cur}^{\rm bub}=2w_{\rm C+L}^{\rm bub}.
\end{equation}
From this equation we find the equation that determines $r_N$:
\begin{equation}\label{rNequation_BUB}
  r_B^4 - 4q^{\rm bub}r_B + 3r^{\rm bub} = 0,
\end{equation}
where
\begin{eqnarray}
\label{qpoly_BUB} q^{\rm bub} = \frac{1}{4}\frac{\sigma_s(x)d}{4\pi(enx)^2f_d(u^{\rm bub})}, \\
\label{rpoly_BUB} r^{\rm bub} = \frac{1}{3}\frac{2d(d-1)\sigma_c(x)}{4\pi(enx)^2f_d(u^{\rm bub})}.
\end{eqnarray}
Notice a difference in signs in Eqs. (\ref{rNequation}) and (\ref{rNequation_BUB}).
The solution to Eq. (\ref{rNequation_BUB}) is
\begin{equation}\label{solution_rN_BUB}
  r_B=\sqrt{\frac{p^{\rm bub}}{2}}\left[1 + \sqrt{q^{\rm bub}\left(\frac{2}{p^{\rm bub}}\right)^{3/2}-1}\right],
\end{equation}
where
\begin{eqnarray}
\nonumber  &&p^{\rm bub}=\left[{q^{\rm bub}}^2 + \sqrt{{q^{\rm bub}}^4-{r^{\rm bub}}^3}\right]^{1/3} \\
\label{p_BUB}  &&+ \left[{q^{\rm bub}}^2-\sqrt{{q^{\rm bub}}^4-{r^{\rm bub}}^3}\right]^{1/3}.
\end{eqnarray}
Equation (\ref{solution_rN_BUB}) is the analytical solution for the bubble radius with curvature tension included.
In case when the curvature contribution is neglected one obtains a simple result,
\begin{equation}\label{solution_rN_noCurvature_BUB}
  r_B=\left[\frac{\sigma_s(x) d}{4\pi(enx)^2f_d(u^{\rm bub})}\right]^{1/3}. \quad(\sigma_c=0)
\end{equation}

\subsection{Chemical equilibrium}
The equation
\begin{equation}\label{var_eq_x_BUB}
  \left.\frac{\partial(w_{\rm tot}^{\rm bub})}{\partial x}\right|_{n,n_{no},r_B,u^{\rm bub}}=0
\end{equation}
is equivalent to the condition of beta equilibrium:
\begin{equation}\label{betaEquilCond_form_BUB}
  \mu_{e}^{\mathrm{bub}}=\mu_{ni}^{\mathrm{bub}} - \mu_{pi}^{\mathrm{bub}},
\end{equation}
where the neutron and the proton chemical potentials are given below by Eqs. (\ref{Muni_BUB}) and (\ref{Mupi_BUB}), respectively.
In the explicit for, this reads
\begin{eqnarray}
 &&  \label{betaEquilCond_BUB}
  \mu_e^{\mathrm{bub}}+(m_p-m_n)c^2 = \\
\nonumber  && -\left.\frac{\partial \varepsilon}{\partial x}\right|_{n} - \frac{1}{(1-u^{\mathrm{bub}})n}\left.\frac{\partial (w_{\rm s}^{\mathrm{bub}}+w_{\rm cur}^{\mathrm{bub}}+w_{\rm C+L}^{\mathrm{bub}})}{\partial x}\right|_{n,n_{no},r_B,u^{\mathrm{bub}}},
\end{eqnarray}
where
\begin{equation}\label{def_mue_BUB}
  \mu_e^{\mathrm{bub}}=\hbar c\left[3\pi^2(1-u^{\mathrm{bub}})xn\right]^{\frac{1}{3}}.
\end{equation}

\subsection{Continuity of neutron chemical potential across the bubble interface}
The extremum defined as
\begin{equation}\label{var_nni_BUB}
  \left.\frac{\partial(w_{\rm tot}^{\mathrm{bub}})}{\partial n_{ni}}\right|_{N_{n},n_{pi},r_B,u^{\mathrm{bub}}}=0
\end{equation}
implies that the energy to add a neutron to the nuclear matter outside the bubble is equal to that in the pure neutron liquid inside the bubble,
\begin{equation}\label{muni_continuity_BUB}
\mu_{ni}^{\mathrm{bub}}=\mu_{no}^{\mathrm{bub}}.
\end{equation}
To obtain the explicit form of this equation we note that fixing of $N_{n}$, $n_{pi}$, $r_B$, $u^{\mathrm{bub}}$ induces the following relations:
\begin{eqnarray}
\label{deltax_deltan_bubbles} && n\delta x + x\delta n=0,\\
\label{deltanni_deltanno_bubbles} &&  (1-u^{\mathrm{bub}})\delta n_{ni} + u^{\mathrm{bub}}\delta n_{no}=0.
\end{eqnarray}
The partial derivative reads
\begin{eqnarray}\label{partial_nnibubbles_initial}
 &&  \left.\frac{\partial}{\partial n_{ni}}\right|_{N_{n},\,n_{pi},\,r_B,\,u^{\mathrm{bub}}}=\left.\frac{\partial}{\partial n}\right|_{x,\,n_{no},\,r_B,\,u^{\mathrm{bub}}} \\
  \nonumber &&  + \frac{\delta x}{\delta n}\left.\frac{\partial}{\partial x}\right|_{n,\,n_{no},\,r_B,\,u^{\mathrm{bub}}} + \frac{\delta n_{no}}{\delta n_{ni}}\left.\frac{\partial}{\partial n_{no}}\right|_{n,\,x,\,r_B,\,u^{\mathrm{bub}}}.
\end{eqnarray}
Using Eqs. (\ref{deltax_deltan_bubbles}) and (\ref{deltanni_deltanno_bubbles}), we find
\begin{eqnarray}\label{partial_nni_BUB}
 &&  \left.\frac{\partial}{\partial n_{ni}}\right|_{N_{n},\,n_{pi},\,r_B,\,u^{\mathrm{bub}}}=\left.\frac{\partial}{\partial n}\right|_{x,\,n_{no},\,r_B,\,u^{\mathrm{bub}}} \\
  \nonumber &&  - \frac{x}{n}\left.\frac{\partial}{\partial x}\right|_{n,\,n_{no},\,r_B,\,u^{\mathrm{bub}}} - \frac{1-u^{\mathrm{bub}}}{u^{\mathrm{bub}}}\left.\frac{\partial}{\partial n_{no}}\right|_{n,\,x,\,r_B,\,u^{\mathrm{bub}}}.
\end{eqnarray}
Application of these partial derivatives yields
\begin{equation}\label{continuityMu_BUB}
  \mu_{ni}^{\mathrm{bub}}=\mu_{no}^{\mathrm{bub}},
\end{equation}
where
\begin{eqnarray}
\nonumber   \mu_{ni}^{\mathrm{bub}}=m_nc^2 + \left(\left.\frac{\partial}{\partial n}\right|_{x} - \frac{x}{n}\left.\frac{\partial}{\partial x}\right|_{n}\right)\left[n\varepsilon(n,x)\right] \\
\label{Muni_BUB}   - \frac{x}{(1-u^{\mathrm{bub}})n} \left.\frac{\partial \left( w_{\rm s}^{\mathrm{bub}} + w_{\rm cur}^{\mathrm{bub}}\right)}{\partial x}\right|_{r_B,\,u^{\mathrm{bub}},},
\end{eqnarray}
and
\begin{equation}\label{Muno_BUB}
  \mu_{no}^{\mathrm{bub}}=\frac{1}{u^{\mathrm{bub}}}\left.\frac{\partial  w_{\mathrm{no}}^{\mathrm{bub}}}{\partial n_{no}}\right|_{u^{\mathrm{bub}}}.
\end{equation}
By combining Eqs. (\ref{betaEquilCond_BUB}), (\ref{betaEquilCond_form_BUB}) and (\ref{Muni_BUB}) we find the proton chemical potential outside the bubble including the surface and Coulomb corrections:
\begin{eqnarray}
\nonumber   &&\mu_{pi}^{\mathrm{bub}}=m_pc^2 + \left(\left.\frac{\partial}{\partial n}\right|_{x} + \frac{1-x}{n}\left.\frac{\partial}{\partial x}\right|_{n}\right)\left[n\varepsilon(n,x)\right] \\
\nonumber  &&+ \frac{1-x}{(1-u^{\mathrm{bub}})n} \left.\frac{\partial  (w_{\rm s}^{\rm bub}+w_{\rm cur}^{\rm bub})}{\partial x}\right|_{r_B,u^{\mathrm{bub}}} \\
\label{Mupi_BUB}&&+ \frac{1}{(1-u^{\mathrm{bub}})n} \left.\frac{\partial  (w_{\rm C+L}^{\rm bub})}{\partial x}\right|_{n,r_B,u^{\mathrm{bub}}}.
\end{eqnarray}

\subsection{Continuity of pressure across the bubble interface}
The extremum
\begin{equation}\label{var_u_BUB}
  \left.\frac{\partial(w_{\rm tot}^{\mathrm{bub}})}{\partial u^{\mathrm{bub}}}\right|_{N_{n},N_{no},x,V_c}=0
\end{equation}
implies that the energy density is minimized by choosing the optimal size of the bubble while keeping fixed the cell size, the number of neutrons, both inside and outside the bubble, and the proton fraction.
As a result, the pressure inside the bubble is equal to that outside the bubble,
\begin{equation}\label{P_continuity_BUB}
P_{i}^{\mathrm{bub}}=P_{o}^{\mathrm{bub}}.
\end{equation}
Fixing $N_{n}$, $N_{no}$, $x$ and $V_c$ induces the following relations:
\begin{eqnarray}
\label{deltau_deltan_bubbles} (1-u^{\mathrm{bub}})\delta n - n\delta u^{\mathrm{bub}}=0,\\
\label{deltanno_deltau_bubbles} n_{no} \delta u^{\mathrm{bub}} + u^{\mathrm{bub}}\delta n_{no}=0.
\end{eqnarray}
The partial derivative reads
\begin{eqnarray}\label{partial_ububbles_initial}
 &&  \left.\frac{\partial}{\partial n_{ni}}\right|_{N_{n},\,n_{pi},\,r_B,\,u^{\mathrm{bub}}}=\left.\frac{\partial}{\partial u^{\mathrm{bub}}}\right|_{n,\,x,\,n_{no},\,V_c} \\
  \nonumber &&  + \frac{\delta n}{\delta u^{\mathrm{bub}}}\left.\frac{\partial}{\partial n}\right|_{n_{no},\,x,\,r_B,\,u^{\mathrm{bub}}} + \frac{\delta n_{no}}{\delta u^{\mathrm{bub}}}\left.\frac{\partial}{\partial n_{no}}\right|_{n,\,x,\,r_B,\,u^{\mathrm{bub}}}.
\end{eqnarray}
Using Eqs. (\ref{deltau_deltan_bubbles}) and (\ref{deltanno_deltau_bubbles}), we find
\begin{eqnarray}\label{partial_ububbles}
 &&  \left.\frac{\partial}{\partial u^{\mathrm{bub}}}\right|_{N_{n},\,N_{no},\,x,\,V_c}=\left.\frac{\partial}{\partial u^{\mathrm{bub}}}\right|_{n,\,x,\,n_{no},\,V_c} \\
  \nonumber &&  + \frac{n}{1-u^{\mathrm{bub}}}\left.\frac{\partial}{\partial n}\right|_{n_{no},\,x,\,r_B,\,u^{\mathrm{bub}}} - \frac{n_{no}}{u^{\mathrm{bub}}}\left.\frac{\partial}{\partial n_{no}}\right|_{n,\,x,\,r_B,\,u^{\mathrm{bub}}}.
\end{eqnarray}
Making use of the relation
\begin{equation}\label{relation_bubbles}
\left.\partial[(1-u^{\mathrm{bub}})n]/\partial u^{\mathrm{bub}}\right|_{N_{n},\,N_{no},\,x,\,V_c}=0,
\end{equation}
we obtain Eq. (\ref{P_continuity_BUB}), where
\begin{eqnarray}
  \label{Pi_BUB} && P_{i}^{\mathrm{bub}} = n^2\left.\frac{\partial \varepsilon}{\partial n}\right|_{x}+ \left(\left.\frac{\partial}{\partial u^{\mathrm{bub}}}\right|_{n,\,x,\,n_{no},\,V_c}\right. \\
  \nonumber &&+ \left.\frac{n}{1-u^{\mathrm{bub}}}\left.\frac{\partial}{\partial n}\right|_{n_{no},\,r_B,\,u^{\mathrm{bub}}}\right)(w_{\rm s}^{\mathrm{bub}}+w_{\rm cur}^{\mathrm{bub}} + w_{\rm C+L}^{\mathrm{bub}}),
\end{eqnarray}
and
\begin{equation}\label{Po_BUB}
  P_{o}^{\mathrm{bub}} = n_{no}^2\frac{\partial \varepsilon(n_{no},0)}{\partial n_{no}}.
\end{equation}
With the help of Eq. (\ref{deltau_deltan}) and the relations
\begin{eqnarray}
&& \left.\partial(u^{\mathrm{bub}}/r_B)/\partial u^{\mathrm{bub}}\right|_{V_c}=(d-1)/r_Bd,\\
&& \left.\partial(u^{\mathrm{bub}}/r_B^2)/\partial u^{\mathrm{bub}}\right|_{V_c}=(d-2)/r_Bd,\\
&& \left.\partial(r_B^2)/\partial u^{\mathrm{bub}}\right|_{V_c}=2r_B^2/u^{\mathrm{bub}}d,
\end{eqnarray}
the derivatives within $P_{i}^{\mathrm{bub}}$ are calculated as following:
\begin{widetext}
\begin{eqnarray}
&& \label{dws_dububbles}\left.\frac{\partial w_{\rm s}^{\mathrm{bub}}}{\partial u}\right|_{N_{n},\,N_{no},\,x,\,V_c}=\frac{d-1}{r_B}\sigma_s(x),\qquad\qquad\qquad \left.\frac{\partial w_{\rm cur}^{\mathrm{bub}}}{\partial u}\right|_{N_{n},\,N_{no},\,x,\,V_c}=-\frac{(d-2)(d-1)}{r_B^2}\sigma_c(x),\\
&& \nonumber\left(\left.\frac{\partial}{\partial u}\right|_{n,\,x,\,n_{no},\,V_c} + \frac{n}{1-u^{\mathrm{bub}}}\left.\frac{\partial}{\partial n}\right|_{n_{no},\,r_B,\,u^{\mathrm{bub}}}\right)w_{\rm C+L}^{\mathrm{bub}}
  =2\pi (e x n r_B)^2 \times\left[\left(\frac{2+d}{d} + \frac{2u^{\mathrm{bub}}}{1-u^{\mathrm{bub}}}\right)f_d(u^{\mathrm{bub}}) + u^{\mathrm{bub}}\partial_{u^{\mathrm{bub}}}f_d(u^{\mathrm{bub}})\right].
\end{eqnarray}
\end{widetext}

\section{Solution of Eqs. (\ref{rNequation}) and (\ref{rNequation_BUB})}
Analytical solution to the quartic Eqs. (\ref{rNequation}) and (\ref{rNequation_BUB}) determining the nuclear and the bubble radii can be obtained as following.
Write the quartic equation in a generic reduced form
\begin{equation}\label{rNeqQuartic}
  X^4-4qX-3r = 0,
\end{equation}
where $X=r_N$, $r>0$ for the nuclei and $X=r_B$, $r<0$ for the bubbles.
The basic idea is to reduce Eq. (\ref{rNeqQuartic}) to a quadratic equation of the form
\begin{equation}\label{Quartic1}
(X^2+m)^2=(m_1X+m_2)^2,
\end{equation}
where $m$, $m_1$ and $m_2$ are to be determined.
Using the elementary result $(X^2+m)^2=X^4+2mX^2+m^2$ and expressing $X^4$ from Eq. (\ref{rNeqQuartic}), yields
\begin{equation}\label{Quartic2}
  (X^2+m)^2=2mX^2+4qX+3r+m^2.
\end{equation}
It is now obvious that the form specified in Eq. (\ref{Quartic1}) can be achieved by choosing $m_1=\sqrt{2m}$ and $m_2=2q/\sqrt{2m}$.
Therefore, $m$ satisfies the cubic equation
\begin{equation}\label{Quartic3}
  m^2+3r=\frac{2q^2}{m}.
\end{equation}
The standard method to solve a cubic equation is well-known \cite{Cardano1545} and consists in representing $m$ as a sum of two new variables,
\begin{equation}\label{def_m}
m=v+w,
\end{equation}
rendering Eq. (\ref{Quartic3}) to the form
\begin{equation}\label{Quartic4}
  v^3+w^3+3(vw+r)(v+w)-2q^2=0.
\end{equation}
The next step is to require that the pair $v$, $w$ satisfy the following conditions
\begin{equation}\label{Quartic5}
  wv=-r,\quad v^3+w^3=2q^2.
\end{equation}
Therefore, $v^3$ and $w^3$ both satisfy quadratic equations with the solutions
\begin{equation}\label{Quartic6}
  v^3=q^2\pm\sqrt{q^4+r^3},\quad w^3=q^2\mp\sqrt{q^4+r^3}.
\end{equation}
Equations (\ref{def_m}) and (\ref{Quartic6}) immediately resolve Eq. (\ref{Quartic3}):
\begin{equation}\label{m_solution}
  m=\left[q^2\pm\sqrt{q^4+r^3}\right]^{\frac{1}{3}}+\left[q^2\mp\sqrt{q^4+r^3}\right]^{\frac{1}{3}},
\end{equation}
while Eq. (\ref{Quartic1}) implies
\begin{equation}\label{X_solution}
  X^2+m=\pm\left(\sqrt{2m}X+\frac{2q}{\sqrt{2m}}\right).
\end{equation}
It is clear that Eq. (\ref{m_solution}) is degenerate with respect to choosing of either $+$ and $-$ in the first and the second term in the right-hand side or, correspondingly, of $-$ and $+$.
Finally, the four linearly independent solutions to Eq. (\ref{rNeqQuartic}) are
\begin{eqnarray}
&& \label{X12}  X_{1,2}=\pm\sqrt{\frac{m}{2}} + \sqrt{\pm\frac{2q}{\sqrt{2m}}-\frac{m}{2}}, \\
&& \label{X34}  X_{3,4}=\pm\sqrt{\frac{m}{2}} - \sqrt{\pm\frac{2q}{\sqrt{2m}}-\frac{m}{2}}.
\end{eqnarray}
The physical solution can be identified by examination of the limiting case of Eq. (\ref{rNeqQuartic}) for $r=0$.
Denote by $X_0$ the solution of the limiting case, that is $X_0=(4q)^{1/3}$.
Assuming $r=0$ in Eq. (\ref{X12}) and (\ref{X34}), we find that the correct value, $X_0$, can be obtained only from Eq. (\ref{X12}) with plus signs:
\begin{equation}\label{X}
  X=\sqrt{\frac{m}{2}} + \sqrt{\frac{2q}{\sqrt{2m}}-\frac{m}{2}}.
\end{equation}

\end{document}